\documentclass[prx,singlecolumn,english,superscriptaddress,floatfix]{revtex4-2}
\usepackage{caption}
\usepackage{graphicx}%
\usepackage{dcolumn}%
\usepackage{bm}
\usepackage{amsmath}

\usepackage{mathtools}
\usepackage{tikz}
\usepackage{subcaption} %
\usepackage{amssymb}
\usetikzlibrary{arrows.meta,calc}
\usepackage{xcolor}
\usepackage{braket}
\usepackage{booktabs}
\usepackage{url}
\usepackage{bbold}
\usepackage{indentfirst}
\usepackage{amsfonts}
\usepackage{float}

\begin{document}

\title{Constructively Embedding Quantum Error-Correcting Codes}%

\author{Yucheng He}
 \email{yucheng.he@usc.edu}
 \affiliation{Ming Hsieh Department of Electrical and Computer Engineering, University of Southern California, Los Angeles, California 90007, USA}%

\author{Todd A. Brun}%
\affiliation{Ming Hsieh Department of Electrical and Computer Engineering, University of Southern California, Los Angeles, California 90007, USA}%

\date{\today}

\begin{abstract}
\noindent

Quantum error-correcting codes must fit physical hardware, not only have good parameters. Surface codes map naturally onto two-dimensional arrays, but many quantum low-density parity-check (qLDPC) codes have Tanner graphs with nonlocal connections. We develop a constructive pipeline that maps a Tanner graph onto a target hardware lattice in three stages: continuous embedding, grid assignment, and path routing. Continuous embeddings reveal the geometric structure of a code, including differences between stabilizer generator sets of the same code. Proxy metrics for footprint, communication overhead, and congestion show how routing geometry, dimension, and scale affect the routed layouts. Ablation studies on the HaPPY code and nine qLDPC codes show that the continuous embedding matters more than the grid assignment. We also compare our approach with Hardware-Aware Layout (HAL) [M. Mathews \emph{et al.}, npj Quantum Inf.\ \textbf{12}, 114 (2026)]. Passing our layouts to HAL's router lowers its hardware overhead for nearly all codes and embeddings tested, which suggests that the two approaches are complementary.

\end{abstract}

\maketitle

\section{Introduction}
\label{secI}
Scalable quantum error correction is not only a coding problem; it is also a problem of layout design. A code with excellent algebraic parameters must still be embedded, routed, and measured in physical hardware. The surface code is a standard example: its checks are local on a two-dimensional lattice~\cite{DennisKitaevLandahlPreskill2002,kitaev2003,fowler2012,GoogleBelowThreshold2025}. However, this locality comes at a cost. Rate-distance bounds for two-dimensional geometrically local codes, such as the Bravyi--Poulin--Terhal bound~\cite{bravyi2010}, constrain the achievable trade-offs between logical information, code distance, and physical-qubit overhead.

Quantum low-density parity-check (qLDPC) codes offer a route beyond this locality-constrained regime. Constructions such as quantum Tanner codes and bivariate bicycle codes use sparse parity checks to improve trade-offs on the coding-theoretic side~\cite{ldpc_citation,PanteleevKalachevGoodQLDPC,LeverrierZemorQuantumTanner,BravyiBB2024}. Yet the Tanner graphs that make these codes mathematically attractive can require interactions between distant parts of a device. Given the Tanner graph of a code and a chosen hardware geometry, the central question is how to constructively design a layout that realizes its connections.

The question is timely because the current experimental platforms are moving beyond strictly planar graphs with nearest-neighbor connections. Neutral-atom arrays, multilayer architectures in the superconducting platform, long-range couplers, and modular architectures offer ways of coupling distant qubits~\cite{Bluvstein2024LogicalProcessor,FoldingBB,HAL2025}. Richer connectivity does not remove geometric constraints, however. Nonlocal interactions still consume spacetime volume, routing capacity, or control bandwidth. A useful method should therefore map a Tanner graph onto a target routing geometry and quantify its architectural cost, analogous to placement and routing in classical very-large-scale integration (VLSI) physical design~\cite{Kahng2011VLSI}.

Several recent works address this co-design task. QECC-Synth formulates layout synthesis for QEC syndrome-extraction circuits on sparse architectures~\cite{Yin2023QECCSynth}. Closest in spirit to our setting, Hardware-Aware Layout (HAL) automates the placement and routing of arbitrary QECC connectivity graphs on multilayer superconducting hardware and uses a hardware-cost model to compare code families~\cite{HAL2025}. At a theoretical level, wire-code constructions relate local subsystem-code overhead to embeddings of an input Tanner graph~\cite{WireCodes2026}. Complementary platform-specific studies optimize qLDPC layouts, routing strategies, or measurement protocols for particular architectures~\cite{Berthusen2025QLDPC2D,FoldingBB,Pecorari2025LongRangeNeutralAtoms,Liu2026EfficientRouting}. Collectively, these works address complementary aspects of the problem: layout synthesis, platform-specific routing, and theoretical embeddings. However, they do not provide a common geometry-aware cost model for comparing different Tanner-graph layouts.

Here we use layout construction to diagnose how well a particular Tanner-graph presentation fits a physical routing geometry. Following the standard VLSI workflow, we split the problem into three stages: continuous embedding, grid assignment, and path routing~\cite{Kahng2011VLSI}. Unlike routing for a one-shot quantum circuit~\cite{LiDingXie2019SABRE,Cowtan2019QubitRouting}, Tanner-graph interactions recur in every syndrome-extraction cycle. We therefore track footprint, communication overhead, and parallel-time lower bound as geometric proxies for recurring architectural costs.

We use simple, analyzable algorithms at each stage: spectral embedding for the continuous embedding stage, the Hungarian algorithm for grid assignment, and shortest-path routing on the target lattice. Grid assignment minimizes the displacement of each vertex from its continuous position, while path routing minimizes individual path lengths.

Applying this pipeline to representative codes shows how routing geometry, spatial dimension, scale, and stabilizer generator sets affect the resulting costs. We also run two ablation studies on the R2 HaPPY code. A \emph{cross-stage} ablation compares the default pipeline with alternatives at each stage: graph-aware continuous embedding methods and path routing have the largest effects on the total cost, while replacing the Hungarian algorithm with greedy assignment changes the result relatively little. A \emph{within-stage} study compares distance-based MDS, Laplacian-based spectral embedding, and force-based spring layout. Repeating the cross-stage ablation on nine qLDPC codes yields the same conclusion for the continuous embedding.

It is interesting to compare the approach in this paper to the recent work on HAL~\cite{HAL2025}. HAL lays out a planar subgraph of the Tanner graph and routes the remaining edges on extra tiers of a multilayer superconducting chip. We compare this to laying out the whole Tanner graph with the first two stages of our pipeline, and scoring the result with HAL's own router and cost metric. We find that the cost decreases in $26$ of the $27$ combinations of code and embedding.

The rest of the paper is organized as follows. Section~\ref{sec:code_embedding} introduces the continuous embedding methods and uses them to visualize several codes. Section~\ref{sec:discrete_embedding} describes scale selection, grid assignment, and path routing, and defines the cost metrics. Section~\ref{sec:results} presents the routed layouts, the ablation studies, and the comparison with HAL, and Sec.~\ref{sec:conclusion} concludes. Algorithmic details are given in the appendices.

\section{Code Embedding}
\label{sec:code_embedding}

\subsection{General Embedding Strategy}
\label{sec:general_strategy}

We represent a syndrome-extraction scheme by a Tanner graph $G_T=(V_T,E_T)$. Its vertices comprise two disjoint sets: data nodes, corresponding to physical data qubits, and check nodes, corresponding to ancillas used to measure stabilizer generators. Each edge encodes a required data-check interaction in a syndrome-extraction round~\cite{Gottesman1997Thesis,fowler2012}. For non-CSS generators, such as those of the HaPPY codes, we place the $X$ and $Z$ parts of each generator on separate check nodes, following the symplectic representation of the stabilizers. A layout on a target lattice $G_\Lambda=(\Lambda,\mathcal E_\Lambda)$ consists of an injective assignment of the vertices in $V_T$ to lattice sites, together with a path on $G_\Lambda$ for every edge in $E_T$.

In the ideal case, every data-check pair joined by a Tanner-graph edge sits on neighboring sites that the hardware can couple directly, as for the surface code; syndrome extraction then requires no additional communication. In general, however, some of these pairs are far apart, and their interaction must be mediated by moving qubits, by long-range couplers, by distributing nonlocal entangled resource states or otherwise routing information along a path. The physical cost of this mediation depends on the platform, but its geometric part can be compared across layouts using the proxies defined in Sec.~\ref{sec:qor}. Finding an optimal layout is NP-hard in general~\cite{GareyJohnson1979}. We therefore solve the problem in separate stages with a heuristic pipeline rather than searching for a globally optimal layout.

The continuous stage places the Tanner-graph vertices at real-valued coordinates. These coordinates are useful on their own for visualizing the structure of a code, and they also serve as the starting point for the discrete stages: we rescale them (scale selection), assign each vertex to a distinct lattice site (grid assignment), and route every edge along a lattice path (path routing).

\subsection{Continuous embedding methods}
\label{sec:continuous_init}

We consider three continuous embedding methods, each emphasizing a different aspect of graph structure: \emph{distance-based} multidimensional scaling (MDS), \emph{Laplacian-based} spectral embedding, and \emph{force-based} spring layout. We use them both to visualize the Tanner graph and as interchangeable initializers whose routed costs can be compared (Sec.~\ref{sec:ablation}).

{\bf MDS} places vertices at Euclidean coordinates $\mathbf{x}_i=p(v_i)\in\mathbb R^d$ to represent pairwise graph distances~\cite{Torgerson1952MDS}. In our setting, $d_{ij}=d_T(i,j)$ is the shortest-path distance between vertices $i$ and $j$ on the Tanner graph. A common measure of distance preservation is the normalized stress
\begin{equation}
\sigma(p) = \sqrt{\frac{\sum_{i<j} \left(d_{ij}-\|\mathbf{x}_i-\mathbf{x}_j\|_2\right)^2}{\sum_{i<j}d_{ij}^2}}.
\end{equation}
Because this measure includes all vertex pairs rather than only connected pairs, MDS emphasizes global distance relationships. The distinction between classical and metric MDS is noted in Appendix~\ref{app:mds}.

{\bf Spectral embedding} starts from the adjacency matrix $A$ of the Tanner graph, with $A_{ij}=1$ if $(i,j)\in E_T$ and $A_{ij}=0$ otherwise. Drawing on spectral graph theory, Laplacian eigenmaps, and spectral community detection~\cite{Fiedler1973AlgebraicConnectivity,BelkinNiyogi2003LaplacianEigenmaps,PhysRevE.community}, it uses the eigenvectors associated with the smallest nonzero eigenvalues of the graph Laplacian. Under orthogonality and normalization constraints (Appendix~\ref{app:spectral}), these minimize the edge-based quadratic form
\begin{equation}
\mathcal Q(p) = \sum_{(i,j)\in E_T}\|\mathbf x_i-\mathbf x_j\|_2^2,
\end{equation}
where $\mathbf x_i=p(v_i)\in\mathbb R^d$ is the coordinate of vertex $v_i$. The Laplacian's zero modes are constant within connected components and carry no relative position information, so we discard them and use the next $d$ eigenvectors. A full eigendecomposition costs $O(n^3)$, but sparse solvers can compute the required low-lying eigenvectors efficiently for the graphs considered here.

The {\bf spring (force-directed) layout} treats Tanner-graph vertices as repelling particles joined by attractive springs along the edges~\cite{Fruchterman1991,Kamada1989}. Its iterative equilibrium balances repulsion between vertices against attraction along connected pairs. The result depends on the force parameters, initialization and iteration schedule, and unlike spectral embedding, it has no closed-form solution (Appendix~\ref{app:spring}).

We use spectral embedding as the default initializer: it makes modular structure easy to see, and on the R2 HaPPY code it gives more stable routed costs across scales than spring layout (Sec.~\ref{sec:ablation}).

\subsection{Code visualization}
\begin{figure}[htbp]
    \centering
    \includegraphics[width=1.0\textwidth]{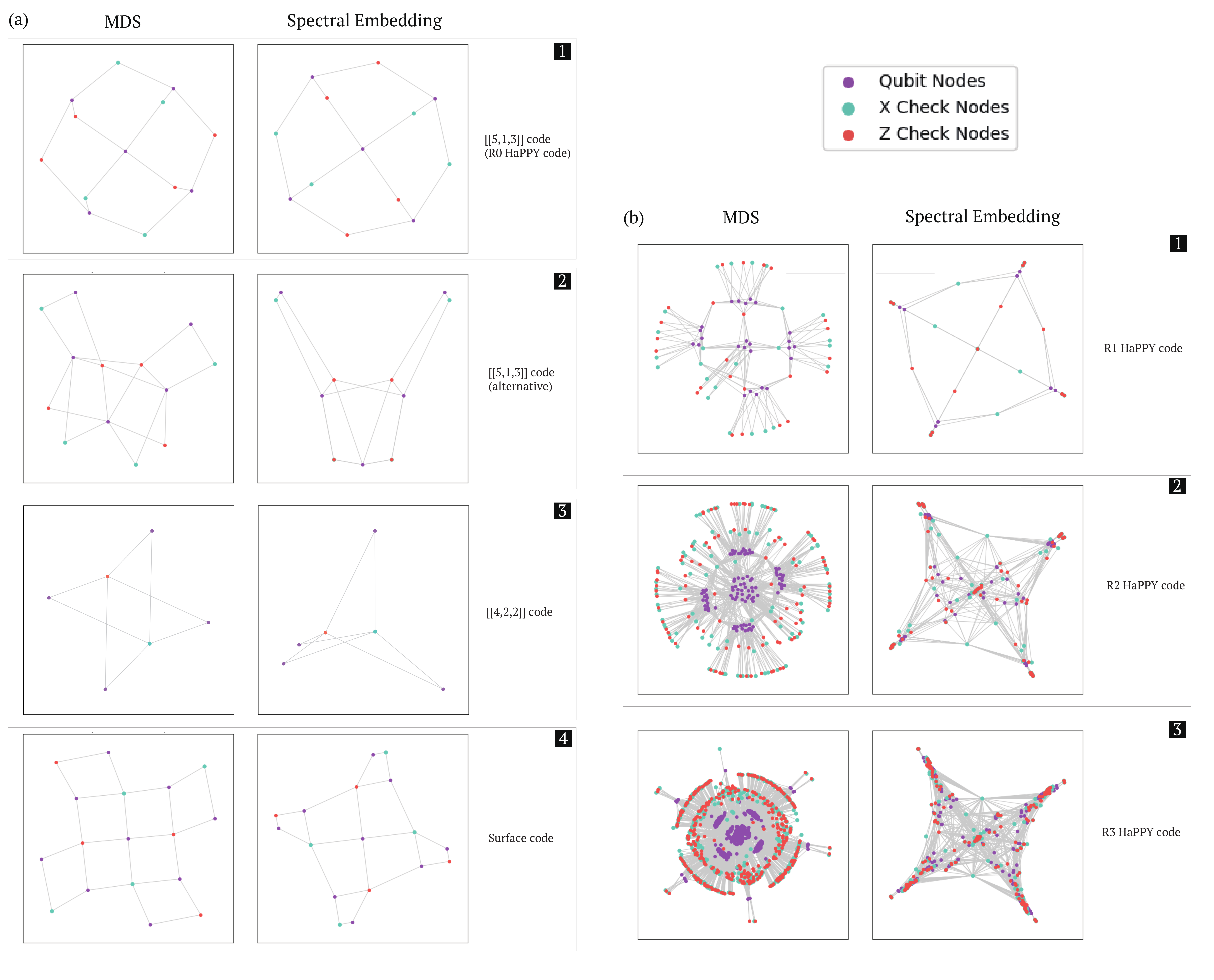}
    \caption{Comparison of 2D embedding techniques for various quantum codes. (a) MDS (left column) versus spectral embedding (right column) for different stabilizer generator sets of the $[[5,1,3]]$ code, the $[[4,2,2]]$ code, and the $[[9,1,3]]$ surface code. (b) Embeddings of the holographic HaPPY code at increasing radii (R0--R3), comparing MDS (left) and spectral embedding (right). Spectral embedding brings out community structure and modularity, while MDS prioritizes global pairwise distances. Node legend: qubits (purple), $X$ checks (turquoise), $Z$ checks (orange).}
    \label{2D}
\end{figure}

\begin{figure}[htbp]
    \centering
    \includegraphics[width=0.9\textwidth]{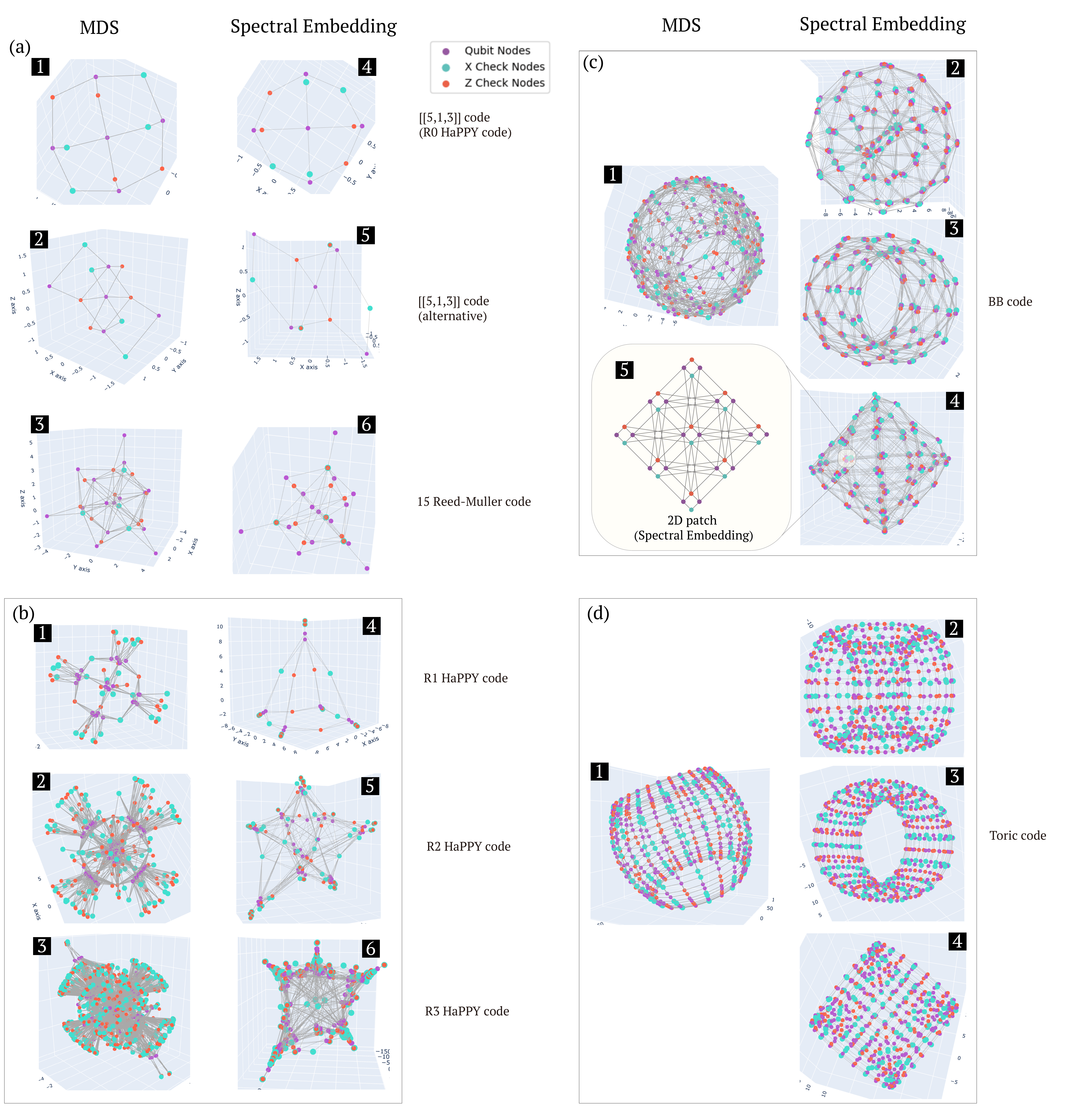}
    \caption{Comparison of 3D embedding techniques and the geometry they reveal. (a--d) Three-dimensional embeddings of the Reed--Muller, $[[5,1,3]]$ (two variants), toric, BB, and HaPPY (R0--R3) codes, comparing MDS (left) with spectral embedding (right). Spectral embedding exposes features such as the pyramidal form of the Reed--Muller code and the concentric layers of the HaPPY code, whereas MDS focuses on global distances. In the spectral embedding of the BB code, panel (c,4), the vertices group into clusters of four: two data qubits and two ancillas, one $X$ check and one $Z$ check. Panel (c,5) displays a 2D patch view derived from the spectral embedding of the BB code, which magnifies part of panel (c,4) to show how these clusters are interwoven.}
    \label{3D}
\end{figure}

Figures~\ref{2D} and~\ref{3D} show 2D and 3D embeddings for a variety of codes: the $[[5,1,3]]$ and $[[4,2,2]]$ codes, surface and toric codes, bivariate bicycle (BB) codes, Reed--Muller codes, and HaPPY codes~\cite{Pastawski2015}. For the HaPPY codes, R$r$ denotes the code of radius $r$: R0 is the $[[5,1,3]]$ code, and R1, R2, and R3 encode one logical qubit in $25$, $95$, and $355$ physical qubits, respectively. For the toric and BB codes, the embeddings recover recognizable toroidal or quasi-local geometry. In the spectral embedding of the BB code, the $288$ vertices of its Tanner graph collapse into $72$ clusters, one for each element of the group $\mathbb{Z}_{12}\times\mathbb{Z}_6$ that defines the code; each cluster contains two data qubits, one $X$ check, and one $Z$ check, reflecting the translation symmetry of the code [Fig.~\ref{3D}(c,4) and (c,5)].

The two Tanner-graph presentations of the $[[5,1,3]]$ code describe the same code, yet their embeddings are visibly different. The same stabilizer code can be written with different stabilizer generator sets, and each set yields a different Tanner-graph presentation with its own check weights, crossings, and apparent locality. Section~\ref{sec:discrete_embedding} turns these embeddings into routed layouts.

\section{Discrete Embedding}
\label{sec:discrete_embedding}

\begin{figure}[htbp]
    \centering
    \includegraphics[width=0.8\textwidth]{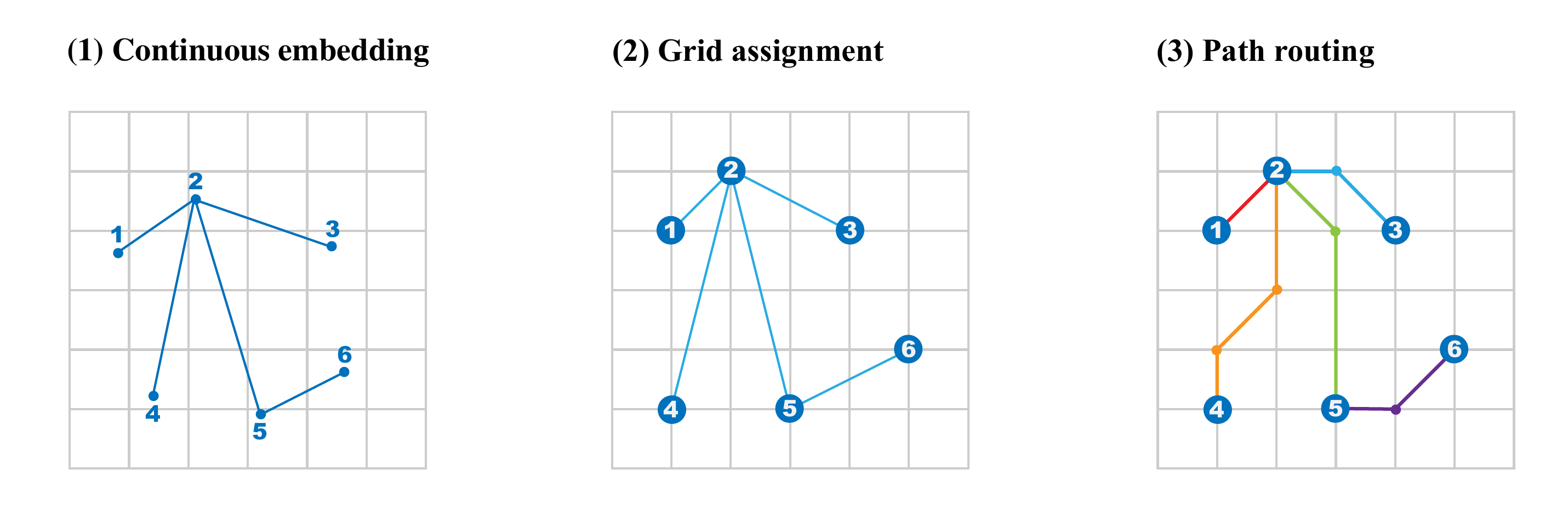}
    \caption{The three-stage flow of code embedding: (1) scale selection and continuous embedding; (2) one-to-one grid assignment; (3) geometry-aware routed lattice.}
    \label{three_step}
\end{figure}

The discrete stage maps the continuous coordinates onto a target lattice with a specified routing geometry. We consider two geometries: rectilinear lattices, which allow only axis-aligned moves and induce the Manhattan distance, and octilinear lattices, which also allow diagonal moves and induce the Chebyshev distance. The three steps of this stage are described below and summarized in Fig.~\ref{three_step}.

\subsection{Scale selection}
\label{subsec:scale_selection}
Let \(p(v)\) denote the continuous coordinates of Tanner-graph vertex \(v\) produced by any of these methods. Their absolute scale is arbitrary. We introduce a scale parameter \(s>0\) to generate scaled coordinates,
\begin{equation}
\hat p_s(v)=s p(v),
\end{equation}
and run grid assignment and routing at each candidate scale. The scale controls how dense the layout is. If \(s\) is too small, many vertices may try to occupy the same lattice sites; grid assignment must displace them from their scaled positions, and more routes must pass through the same sites. If \(s\) is too large, the layout spreads out and both the footprint and the communication overhead grow.

Because the layout is discrete, \(J(s)\) (Eq.~\eqref{eq:scale_obj_new}) is piecewise constant in \(s\): it stays fixed while the assignment and the routes stay fixed, and it jumps when a small change in \(s\) moves a vertex to a different site. Its derivative is therefore zero or undefined, so we select \(s\) with a discrete sweep or a coarse-to-fine search rather than by gradient-based optimization.

\subsection{Grid Assignment}
\label{subsec:grid_assignment}
Let \(\Lambda\) denote the hardware lattice and \(\hat p_s(v)\) the scaled position of vertex \(v\). Grid assignment moves each vertex to a lattice site such that no two vertices share a site and the total displacement from the scaled positions is as small as possible. Displacement is measured in the norm matched to the routing geometry,
\begin{equation}
\label{eqn:Euclidean1inf}
\|\cdot\|_* =
\begin{cases}
\|\cdot\|_1, & \text{rectilinear geometry}, \\
\|\cdot\|_\infty, & \text{octilinear geometry}.
\end{cases}
\end{equation}

Rather than considering every site of \(\Lambda\), we only consider sites near the scaled positions. For each vertex \(v\), the candidate set
\begin{equation}
\Gamma_R(v)=\{\ell\in\Lambda:\ \|\hat p_s(v)-\ell\|_*\le R\}
\end{equation}
contains the sites within distance \(R\) of \(\hat p_s(v)\), and \(\mathcal S=\bigcup_{v\in V_T}\Gamma_R(v)\) is the set of all candidate sites.
Since each \(\Gamma_R(v)\) contains a fixed number of sites, \(|\mathcal S|\) grows at most linearly with \(|V_T|\), independently of the size of the full lattice.

We solve for the assignment \(\tilde p:V_T\to\mathcal S\) as a linear assignment problem (LAP),
\begin{equation}
\begin{aligned}
\min_{\mathbf X}\quad & \sum_{v\in V_T}\sum_{\ell\in\mathcal S}c_{v\ell}x_{v\ell},\\
\text{subject to}\quad & \sum_{\ell\in\mathcal S}x_{v\ell}=1 \quad \forall v\in V_T,\\
& \sum_{v\in V_T}x_{v\ell}\leq 1 \quad \forall\ell\in\mathcal S,\\
& x_{v\ell}\in\{0,1\} \quad \forall v\in V_T,\ \forall\ell\in\mathcal S,
\end{aligned}
\label{eq:lap}
\end{equation}
where \(x_{v\ell}=1\) if vertex \(v\) is assigned to site \(\ell\), and \(c_{v\ell}=\|\hat p_s(v)-\ell\|_*\) is its displacement. The two constraints state that every vertex receives exactly one site and that no site holds more than one vertex. We solve the LAP exactly with the Hungarian algorithm. Implementation details, including the rounding of scaled positions, tie-breaking, and the feasibility of the LAP, are given in Appendix~\ref{app:candidate_sites}.

\subsection{Path Routing}
\label{subsec:path_routing}

Path routing connects the two endpoints of each Tanner-graph edge by a path of allowed lattice moves. By default, we use shortest-path routing, which minimizes the length of every routed edge and hence the communication term of the total cost. It also tends to keep the footprint small, but it does not account for congestion. Congestion-aware routing is used only in the ablation study (see Sec.~\ref{sec:ablation} and Appendix~\ref{app:astar}).

In a $d$-dimensional rectilinear lattice, each step changes exactly one coordinate by one unit. The allowed moves are the axis-aligned directions $\{\pm e_m \mid m=1,\dots,d\}$, and the shortest-path distance is the Manhattan distance $\|x-y\|_1$.

An octilinear lattice also allows diagonal moves, in which every coordinate may change by up to one unit at once: in 2D this adds the four $45^\circ$ diagonals, and in 3D also the face and body diagonals. The shortest-path distance then becomes the Chebyshev distance $\|x-y\|_\infty$. Every allowed move, diagonal or not, costs one unit regardless of its Euclidean length.

Our router constructs canonical shortest paths directly from the differences of the endpoint coordinates. For rectilinear layouts it traverses the coordinate axes in a fixed order; for octilinear layouts it first takes diagonal steps that shrink all nonzero offsets at once and then completes the remaining axis-aligned offsets.

\subsection{Performance Metrics}
\label{sec:qor}
We evaluate each layout by three different costs: communication, congestion, and footprint. All of them are computed from the assignment \(\tilde p\) obtained in Sec.~\ref{subsec:grid_assignment} and the set of routed paths \(\Pi=\{\pi_{uv}\}_{(u,v)\in E_T}\). We write \(d_{\mathrm{grid}}(x,y)\) for the shortest-path distance between lattice sites \(x\) and \(y\), i.e., the Manhattan distance on rectilinear lattices and the Chebyshev distance on octilinear ones. Appendix~\ref{app:metric_limitations} discusses what these proxies capture and what they leave out.

\begin{enumerate}
\item \textbf{Communication overhead.}
For each edge $(u,v)\in E_T$, the router produces a path $\pi_{uv}$ between $\tilde p(u)$ and $\tilde p(v)$, and we write $\operatorname{len}(\pi_{uv})$ for its number of lattice steps. We define the SWAP upper-bound proxy
\begin{equation}
\label{eq:UB}
\mathrm{UB} \;=\; 2\sum_{(u,v)\in E_T}\operatorname{len}(\pi_{uv}),
\end{equation}
where the factor of two accounts for a round-trip data interaction. We also track a SWAP lower-bound proxy based only on endpoint distances,
\begin{equation}
\label{eq:LB}
\mathrm{LB} \;=\; 2\sum_{(u,v)\in E_T}\big(d_{\mathrm{grid}}(\tilde p(u),\tilde p(v))-1\big),
\end{equation}
and the total endpoint-to-endpoint grid distance (denoted by $W$ in the results)
\begin{equation}
\label{eq:W}
W_{\mathrm{grid}}=\sum_{(u,v)\in E_T} d_{\mathrm{grid}}(\tilde p(u),\tilde p(v)).
\end{equation}
With shortest-path routing, every path has the minimum possible length, $\operatorname{len}(\pi_{uv}) = d_{\mathrm{grid}}(\tilde p(u),\tilde p(v))$, so $\mathrm{UB} = 2W_{\mathrm{grid}} = \mathrm{LB}+2|E_T|$.

\item \textbf{Congestion.} We quantify the congestion of the routed layout by the parallel-time lower bound ($\mathrm{ParLB}$). Let $\mathcal E_\Lambda$ denote the set of lattice edges. The maximum loads on a lattice edge and on a lattice node are
\begin{align}
\mathrm{EdgeLB} &= \max_{e_\Lambda\in\mathcal E_\Lambda}
\left|\{(u,v)\in E_T: e_\Lambda\in \pi_{uv}\}\right|,\\
\mathrm{NodeLB} &= \max_{x\in\Lambda}
\left|\{(u,v)\in E_T: x\in \pi_{uv}\}\right|.
\end{align}
We then use
\begin{equation}
\label{eq:ParLB}
\mathrm{ParLB}=2\max\{\mathrm{EdgeLB},\mathrm{NodeLB}\}
\end{equation}
as a lower-bound proxy for serialization bottlenecks in a parallel syndrome-extraction round. Below, we refer to $\mathrm{ParLB}$ simply as the congestion.

\item \textbf{Footprint ($F$).}
We count the distinct lattice sites used by placed vertices or routed paths:
\begin{equation}
\label{eq:footprint}
F\;=\;\left|\, \{\tilde p(u):u\in V_T\}\ \cup\ \bigcup_{(u,v)\in E_T}\pi_{uv}\,\right|.
\end{equation}
This is a spatial-resource proxy. Depending on the hardware, it can correspond to qubits, couplers, or transport relays, not necessarily only to physical data qubits.

\end{enumerate}

We combine these quantities into a single total cost,
\begin{equation}
J(s) = w_F F(s) + w_{\mathrm{UB}} \, \mathrm{UB}(s) + w_{\mathrm{Par}} \, \mathrm{ParLB}(s),
\label{eq:scale_obj_new}
\end{equation}
which we use to select the scale and to compare layouts. Every term is evaluated for a given scale $s$. We choose the weights based on a cost model for the given hardware; in the examples below, we use $(w_F,w_{\mathrm{UB}},w_{\mathrm{Par}})=(1,0.02,0.5)$. $\mathrm{UB}$ is much larger than the other two terms, hence its small weight, and the weight on $\mathrm{ParLB}$ makes $J(s)$ respond to congestion.

\begin{figure}[htbp]
    \centering
    \includegraphics[width=0.8\textwidth]{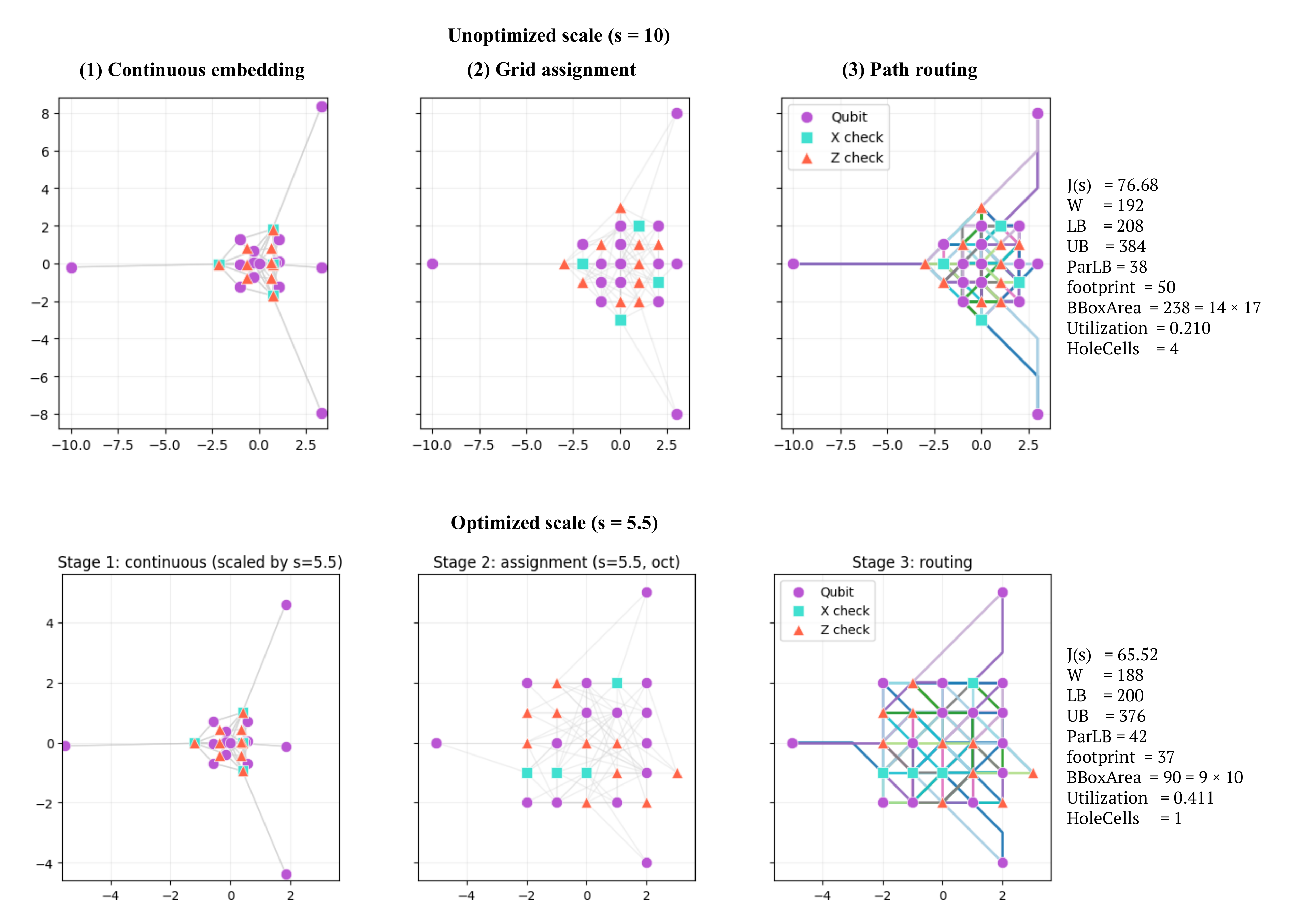}
    \caption{Example three-stage layout for the $[[15,1,3]]$ Reed--Muller code: continuous embedding, grid assignment, and routed edge paths, for two values of the scale $s$.}
    \label{demo}
\end{figure}

Figure~\ref{demo} illustrates the full pipeline on the $[[15,1,3]]$ quantum Reed--Muller code, at $s=10$ and at the selected scale $s=5.5$. Each row of figures goes from continuous embedding through grid assignment to the routed paths on which the metrics are computed. In both rows, a few qubits stay far from the rest of the layout. Because $s$ rescales the whole embedding, they remain somewhat displaced from the main body of the layout, but compressing them further would actually crowd more routes through the sites near the center and increase $J(s)$.

\section{Results and Analysis}
\label{sec:results}

\begin{figure}[htbp]
    \centering
    \includegraphics[width=0.97\textwidth]{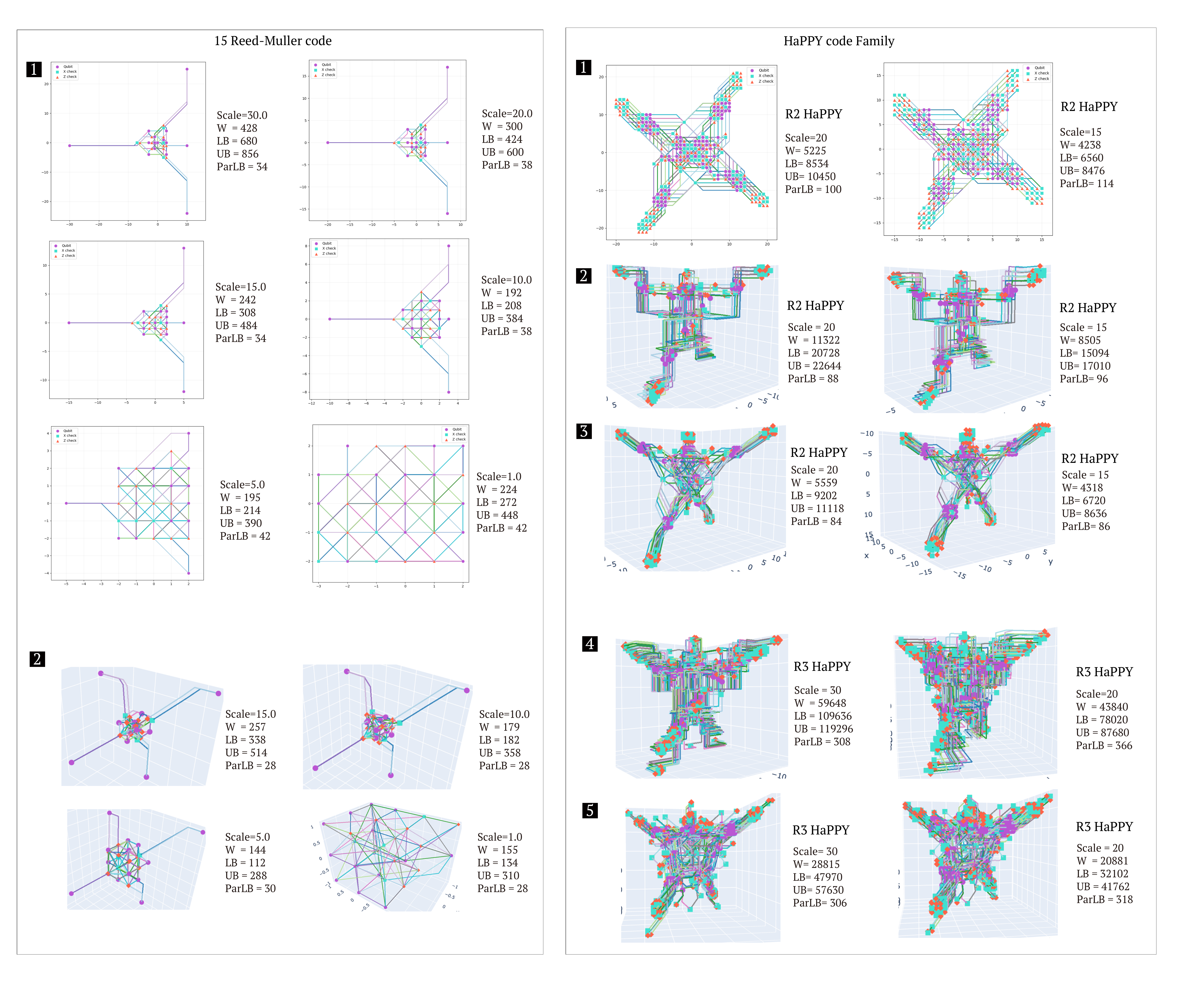}
    \caption{Effect of the scale parameter $s$ on layout quality. The left column shows octilinear layouts of the Reed--Muller code, with the 2D cases in group 1 and the 3D cases in group 2. The right column shows HaPPY-code layouts: octilinear in rows 1, 3, and 5, and rectilinear in rows 2 and 4.}
    \label{assort}
\end{figure}

We apply the pipeline to the HaPPY codes R0--R3, the $[[144,12,12]]$ BB code (the gross code of Ref.~\cite{BravyiBB2024}), the $[[288,2,12]]$ toric code, the $[[9,1,3]]$ surface code, and the $[[15,1,3]]$ Reed--Muller code. Representative routed layouts are shown in Figs.~\ref{rectilinear}, \ref{octilinear}, and \ref{assort}. At the end of Sec.~\ref{sec:ablation} and in Sec.~\ref{sec:hal}, we apply the pipeline to radial, quantum Tanner, and further BB codes from Ref.~\cite{HAL2025}. In Sec.~\ref{sec:hal}, we compare our placement with HAL's placement under its hardware cost model.

\subsection{Choice of Stabilizer Generator Sets, Geometry, and Dimensionality}
\label{sec:interplay}

\begin{figure}[htbp]
    \centering   \includegraphics[width=0.85\textwidth,height=0.85\textheight,keepaspectratio]{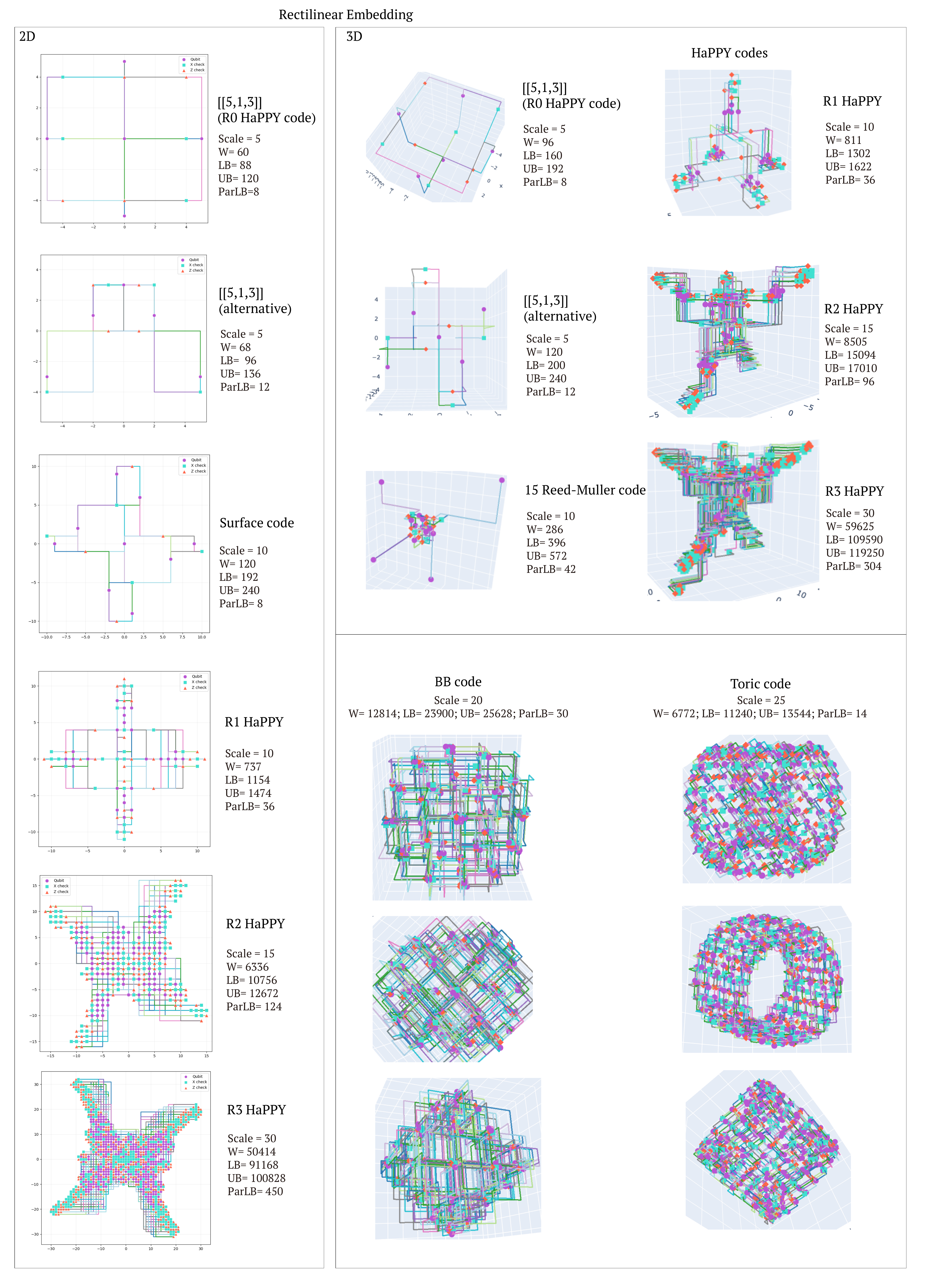}
    \caption{Rectilinear embeddings for representative QECCs. The left column shows 2D layouts for variants of the $[[5,1,3]]$ code, the surface code, and the HaPPY codes R0--R3. The right column shows the corresponding 3D layouts, along with Reed--Muller, BB, and toric code examples. Nodes represent qubits, $X$ checks, and $Z$ checks; panels report \textbf{W}, \textbf{LB}/\textbf{UB}, and \textbf{ParLB}.}
    \label{rectilinear}
\end{figure}

\begin{figure}[htbp]
    \centering    \includegraphics[width=0.93\textwidth,height=0.85\textheight,keepaspectratio]{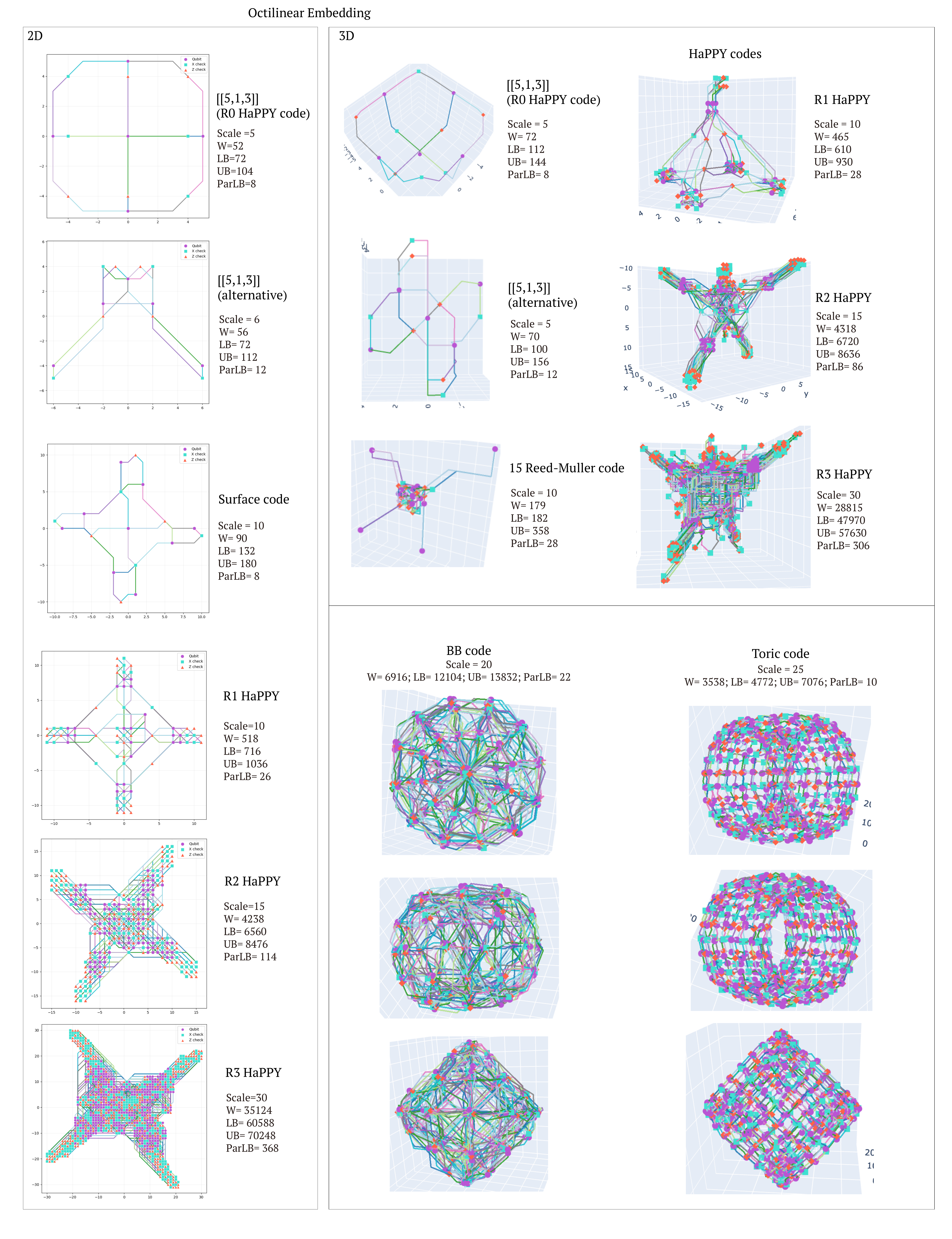}
    \caption{Octilinear embeddings for the surface, HaPPY, Reed--Muller, toric, and BB codes. The left column shows the 2D implementations, and the right column shows their 3D counterparts. The node legend and metrics are as in Fig.~\ref{rectilinear}.}
    \label{octilinear}
\end{figure}

As shown in Figs.~\ref{rectilinear} and \ref{octilinear}, the cost metrics are systematically reduced in switching from rectilinear to octilinear embedding, due to the availability of diagonal moves. It lowers the total grid distance $W$ by about 25--33\% in the 2D examples shown. For the surface code, $W$ goes from $120$ to $90$; for the R1 HaPPY code, from $737$ to $518$; and for the R2 HaPPY code, from $6336$ to $4238$.

In 3D, the reduction is larger still, approaching a factor of two. For the R2 HaPPY code at scale $s=15$, the 3D rectilinear layout has $W=8505$ and $\mathrm{ParLB}=96$, whereas the 3D octilinear layout has $W=4318$ and $\mathrm{ParLB}=86$. For the BB code at $s=20$, the numbers improve from $W=12814$, $\mathrm{ParLB}=30$ (rectilinear) to $W=6916$, $\mathrm{ParLB}=22$ (octilinear).

The effect of going from 2D to 3D is more nuanced. For the R2 HaPPY code in rectilinear geometry, going from 2D to 3D increases $W$ from $6336$ to $8505$, though it lowers $\mathrm{ParLB}$ from $124$ to $96$. In octilinear geometry the 2D and 3D R2 layouts have comparable $W$ ($4238$ vs.\ $4318$), but 3D lowers $\mathrm{ParLB}$ from $114$ to $86$. For the larger R3 HaPPY code, 3D octilinear routing improves both metrics: $W$ drops from $35124$ to $28815$ and $\mathrm{ParLB}$ from $368$ to $306$. In these HaPPY examples, adding the third dimension reliably relieves congestion.

The Tanner-graph presentation is a separate influence on performance. The two $[[5,1,3]]$ examples share the same code parameters but have different Tanner graphs. With 2D rectilinear routing, the R0 HaPPY Tanner-graph presentation gives $W=60$ and $\mathrm{ParLB}=8$, while the alternative Tanner-graph presentation gives $W=68$ and $\mathrm{ParLB}=12$; with octilinear routing, the values are $W=52$, $\mathrm{ParLB}=8$ versus $W=56$, $\mathrm{ParLB}=12$, and the gap persists in 3D. The layout cost therefore depends on the specific Tanner graph, not only on the abstract code.

\subsection{Effect of the Scaling Factor}
\label{sec:scale_impact}

The scale $s$ governs how dense the discrete layout is, which creates a trade-off between $W$ and congestion (Fig.~\ref{assort}). For the 2D octilinear Reed--Muller layout, compressing from $s=30$ to $s=10$ cuts $W$ from $428$ to $192$, while $\mathrm{ParLB}$ shifts only slightly, from $34$ to $38$. Further compression is counterproductive. At $s=5$ we obtain $W=195$, and at $s=1$ the value rebounds to $W=224$. This is the over-compression regime: too many scaled positions round to the same sites, grid assignment must displace them, and edges become longer.

The 3D Reed--Muller layout tolerates compression better. Going from $s=15$ to $s=5$ lowers $W$ from $257$ to $144$, with $\mathrm{ParLB}$ stable at $28$--$30$, and even $s=1$ gives only a mild rebound ($W=155$). The extra room in 3D pushes the onset of over-compression to smaller scales.

The larger HaPPY codes exhibit the same behavior. For the 2D octilinear R2 layout, shrinking $s$ from $20$ to $15$ lowers $W$ from $5225$ to $4238$, but drives $\mathrm{ParLB}$ up from $100$ to $114$. In 3D, the same compression lowers $W$ from $5559$ to $4318$, with $\mathrm{ParLB}$ essentially flat ($84$ to $86$). For R3, compressing the 3D octilinear layout from $s=30$ to $s=20$ brings $W$ down from $28815$ to $20881$, while $\mathrm{ParLB}$ increases slightly from $306$ to $318$. So $s$ must be chosen after routing: denser layouts reduce $W$ up to a point, beyond which collisions and congestion dominate.

\begin{figure}[t]
  \centering
  \includegraphics[width=0.95\textwidth]{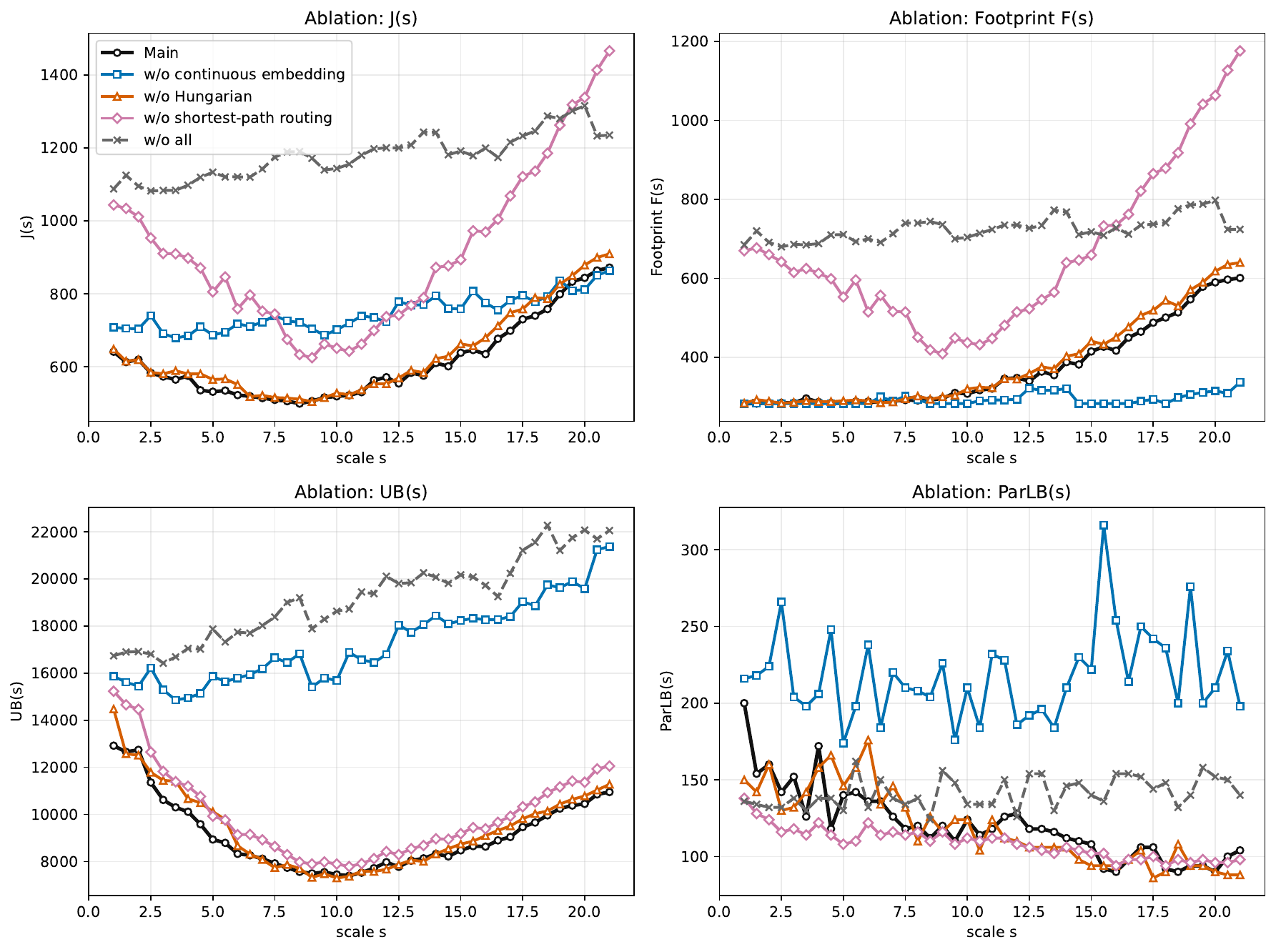}
  \caption{Cross-stage ablation of the default pipeline with weights $(w_F,w_{\mathrm{UB}},w_{\mathrm{Par}}) = (1,0.02,0.5)$. Main = spectral embedding + Hungarian algorithm + shortest-path routing. The three single-stage ablations replace continuous embedding, grid assignment, and path routing, respectively; ``w/o all'' combines all three replacements.}
  \label{fig:ablation-J}
\end{figure}

\subsection{Ablation Study and Baseline Comparisons}
\label{sec:ablation}

\begin{figure}[htbp]
    \centering
    \includegraphics[width=1.0\textwidth]{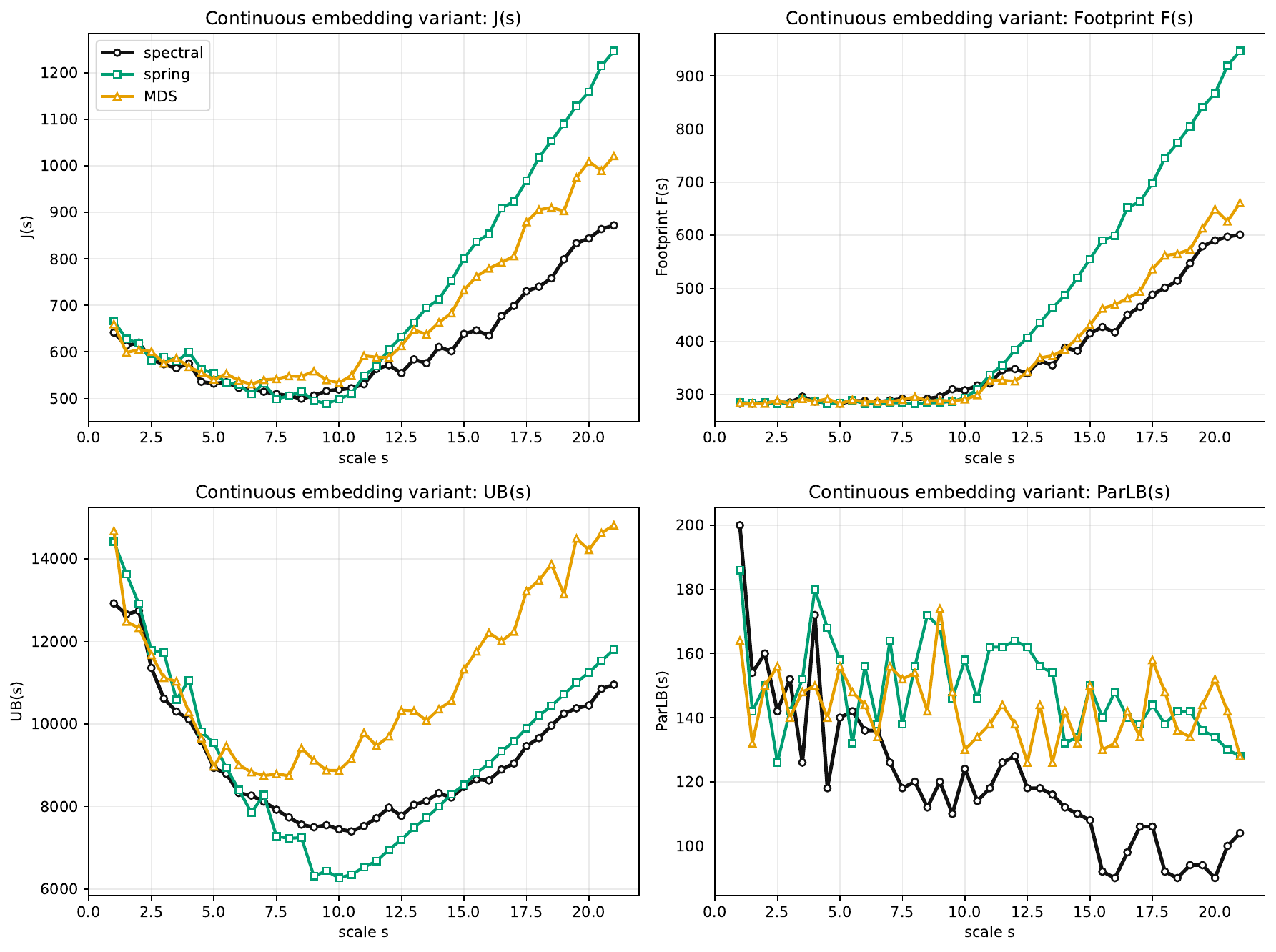}
    \caption{Within-stage substitution at the continuous embedding stage for the R2 HaPPY code, keeping grid assignment and routing fixed. Spectral embedding (black), spring layout (teal), and MDS (orange) are compared on four metrics as functions of scale $s$.}
    \label{fig:ablation-within-stage}
\end{figure}

We organize the ablation study along two axes. The \emph{cross-stage} ablation (Fig.~\ref{fig:ablation-J}) isolates the contribution of each stage by replacing it with a deliberately weakened baseline. The \emph{within-stage} substitution (Fig.~\ref{fig:ablation-within-stage}) keeps assignment and routing fixed and swaps the continuous initializer among the legitimate alternatives (MDS, spectral, spring) to see how the choice of embedding algorithm affects the final cost.

The cross-stage ablation uses the R2 HaPPY code, with the default pipeline
\[
\text{Main} = \text{spectral embedding} + \text{Hungarian algorithm} + \text{shortest-path routing}.
\]
We compare it to three single-stage replacements---type-stratified initialization (a graph-blind initializer that uses only vertex type, to test whether graph awareness is needed; Appendix~\ref{app:type_stratified}), greedy assignment, and congestion-aware A* routing---and against ``w/o all,'' which combines all three. The implementation details are in Appendix~\ref{app:baselines}.

Figure~\ref{fig:ablation-J} shows that, under the weights used here, graph-aware continuous embedding and the choice of router have the largest effects. Without continuous embedding, communication overhead and congestion increase sharply: the $\mathrm{UB}(s)$ curve for the type-stratified initializer stays far above the main curve over essentially the whole scale range, and its $\mathrm{ParLB}(s)$ is often about twice as high, with large spikes at some scales.

Replacing the Hungarian algorithm with greedy assignment has only a small effect. The ``w/o Hungarian algorithm'' curve closely tracks the main curve in $J(s)$, $F(s)$, and $\mathrm{UB}(s)$. Once the continuous embedding is good, the remaining assignment problem is easy enough for a local greedy rule to be competitive.

Replacing shortest-path routing with congestion-aware A* shows a different trade-off. A* reduces local hotspots, giving a lower $\mathrm{ParLB}(s)$ curve, most clearly at small scales. However, it does so through detours, which inflate the footprint at every scale, raise communication overhead slightly, and push $J(s)$ up sharply. With the weights used here, shortest-path routing is the better match to the total cost. The combined ``w/o all'' baseline is poor across the sweep: high $J(s)$, large footprint, and large $\mathrm{UB}(s)$.

Within-stage substitution (Fig.~\ref{fig:ablation-within-stage}) compares the default spectral embedding algorithm to spring layout and MDS, keeping the Hungarian algorithm and shortest-path routing fixed.

The spring variant (force-based) has slightly lower $J(s)$ and noticeably lower $\mathrm{UB}(s)$ than the spectral embedding in an intermediate-scale window $s \in [7, 12]$: its minimum $\mathrm{UB}$ is about $6.3 \times 10^3$, compared to about $7.4 \times 10^3$ for spectral embedding. Outside this window it degrades quickly. For $s \gtrsim 14$ its footprint $F(s)$ and $J(s)$ exceed the spectral values by tens of percent, and the across-scale standard deviation of its $J(s)$ is roughly twice as large.

For congestion, the ordering reverses: the spectral pipeline has lower $\mathrm{ParLB}(s)$ over most of the range. A plausible explanation is that tighter clusters shorten many Tanner-graph edges but concentrate their routed paths into narrow channels.

The MDS variant (distance-based) has higher $\mathrm{UB}(s)$ than spectral embedding across nearly the full scale range and no clear advantage for the other metrics. MDS emphasizes global pairwise distances, while routed communication costs are incurred on Tanner-graph edges. For this code, then, no single embedding is best on every metric at every scale.

We repeat the cross-stage ablation on the nine qLDPC codes of Sec.~\ref{sec:hal} to test whether its conclusions depend on the use of the HaPPY code (Table~\ref{tab:ablation-qldpc}). These codes include the $[[144,12,12]]$ BB code. Removing the continuous embedding raises the minimum total cost by $8$--$24\%$ for every code. Replacing the Hungarian algorithm by greedy assignment changes it by at most $4\%$. Congestion-aware A* routing lowers $\mathrm{ParLB}$ for eight of the nine codes, but its detours raise the minimum total cost by $2$--$36\%$. For the radial $[[32,2]]$ code this increase is only $2\%$, smaller than the effect of greedy assignment on the same code. Replacing all three stages raises the minimum total cost by a factor of $1.4$--$2.3$.

\begin{table}[t]
\centering
\caption{Cross-stage ablation on qLDPC codes (2D octilinear lattice, weights $(w_F,w_{\mathrm{UB}},w_{\mathrm{Par}})=(1,0.02,0.5)$). For each pipeline we take the minimum of $J(s)$ over $13$ scales $s\in[0.3,1.5]\sqrt{|V_T|}$. Columns 4--7 give this minimum relative to Main. The last column gives $\mathrm{ParLB}$ of the A* variant relative to Main, both at their minimum-$J$ scales.}
\label{tab:ablation-qldpc}
\begin{tabular}{lrrccccc}
\toprule
Code & $|V_T|$ & $J_{\min}$ (Main) & w/o cont.\ emb. & w/o Hungarian & w/o shortest path & w/o all & $\mathrm{ParLB}$, A*/Main \\
\midrule
BB $[[18,4,4]]$ & 36 & 63 & 1.08 & 0.98 & 1.08 & 1.36 & 0.81 \\
BB $[[72,8,8]]$ & 144 & 238 & 1.16 & 1.01 & 1.34 & 2.18 & 0.63 \\
BB $[[144,12,12]]$ & 288 & 453 & 1.24 & 1.03 & 1.26 & 2.27 & 0.73 \\
Radial $[[24,2]]$ & 48 & 65 & 1.15 & 0.98 & 1.03 & 1.40 & 0.80 \\
Radial $[[32,2]]$ & 64 & 84 & 1.20 & 0.96 & 1.02 & 1.52 & 0.80 \\
Radial $[[54,8]]$ & 108 & 185 & 1.08 & 1.00 & 1.36 & 2.00 & 0.69 \\
qTanner $[[36,8,3]]$ & 68 & 98 & 1.21 & 1.02 & 1.11 & 1.91 & 1.08 \\
qTanner $n{=}54$ & 102 & 159 & 1.14 & 0.99 & 1.20 & 2.12 & 0.83 \\
qTanner $n{=}72$ & 136 & 203 & 1.23 & 1.03 & 1.07 & 2.27 & 0.74 \\
\bottomrule
\end{tabular}
\end{table}

\subsection{Placement for a Hardware-Specific Router}
\label{sec:hal}

The comparisons above use only geometric costs. To test our placement under a hardware cost model, we use HAL~\cite{HAL2025}, which places and routes Tanner graphs on multilayer superconducting hardware. HAL measures the cost of a layout by its hardware complexity $C_{\mathrm{hw}}$, which combines the number of routing tiers, the average coupler length, the number of bump-bond transitions per coupler, and the number of through-silicon vias per edge. HAL expresses coupler lengths in units of the shortest coupler on the first tier. A surface-code layout has $C_{\mathrm{hw}}=1$. In its default mode, HAL extracts a planar subgraph of the Tanner graph, places it on the first tier with a Kamada--Kawai layout, and routes the remaining edges on higher tiers. HAL can also take a placement of the first tier as an input. In what follows, ``our placement'' refers to the output of our continuous embedding and Hungarian grid assignment on a rectilinear lattice. We fix the scale at $s=0.5\sqrt{|V_T|}$ for every code. We then pass our placement to HAL and apply its router and calculate its hardware complexity score. Details are given in Appendix~\ref{app:hal}.

We used nine codes, three each from the radial, quantum Tanner, and BB families studied in Ref.~\cite{HAL2025}. The Tanner graphs of these codes range from $36$ to $288$ vertices. HAL's default placement is randomized, so we ran it with $10$ seeds for the five smallest codes, $6$ for the next three, and $3$ for the $[[144,12,12]]$ BB code. The spring layout algorithm is also randomized and was run with the same seeds. Spectral embedding and MDS are deterministic and give one layout per code. For a stronger baseline, we also ran HAL with its Kamada--Kawai, spectral, and spring initial layouts, and with its toric placement for the BB codes.

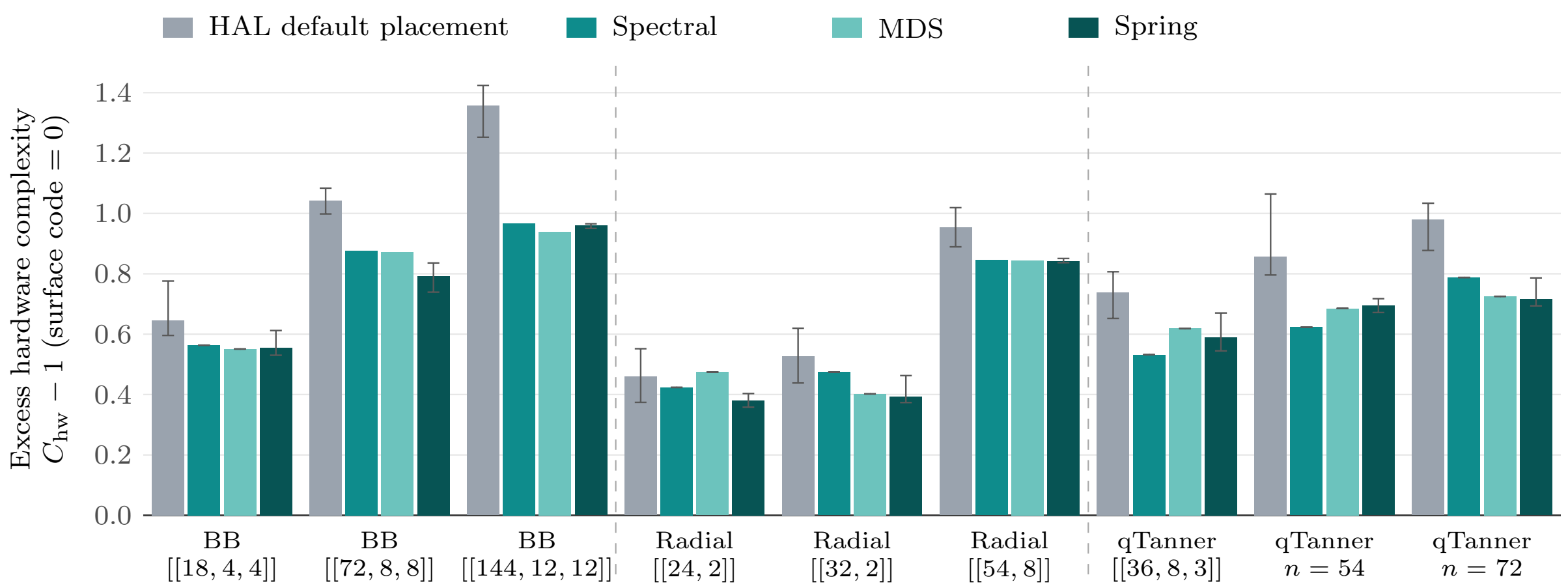
\begin{figure}[t]
  \centering
  \definecolor{halgray}{HTML}{9AA3AE}\definecolor{tspec}{HTML}{0E8C8C}\definecolor{tmds}{HTML}{6CC3BD}\definecolor{tspring}{HTML}{075454}\definecolor{gridgray}{HTML}{E6E6E6}
\begin{tikzpicture}[font=\footnotesize]
\draw[gridgray,line width=0.4pt] (0,0.000) -- (14.400,0.000);
\node[left,text=black!70] at (0,0.000) {0.0};
\draw[gridgray,line width=0.4pt] (0,0.613) -- (14.400,0.613);
\node[left,text=black!70] at (0,0.613) {0.2};
\draw[gridgray,line width=0.4pt] (0,1.227) -- (14.400,1.227);
\node[left,text=black!70] at (0,1.227) {0.4};
\draw[gridgray,line width=0.4pt] (0,1.840) -- (14.400,1.840);
\node[left,text=black!70] at (0,1.840) {0.6};
\draw[gridgray,line width=0.4pt] (0,2.453) -- (14.400,2.453);
\node[left,text=black!70] at (0,2.453) {0.8};
\draw[gridgray,line width=0.4pt] (0,3.067) -- (14.400,3.067);
\node[left,text=black!70] at (0,3.067) {1.0};
\draw[gridgray,line width=0.4pt] (0,3.680) -- (14.400,3.680);
\node[left,text=black!70] at (0,3.680) {1.2};
\draw[gridgray,line width=0.4pt] (0,4.293) -- (14.400,4.293);
\node[left,text=black!70] at (0,4.293) {1.4};
\draw[black!80,line width=0.5pt] (0,0) -- (14.400,0);
\draw[black!35,dashed,line width=0.4pt] (4.800,-0.6) -- (4.800,4.600);
\draw[black!35,dashed,line width=0.4pt] (9.600,-0.6) -- (9.600,4.600);
\node[rotate=90,align=center] at (-1.05,2.300) {Excess hardware complexity\\ $C_{\mathrm{hw}}-1$ (surface code $=0$)};
\fill[halgray] (0.088,0) rectangle (0.418,1.979);
\draw[black!65,line width=0.45pt] (0.253,1.827) -- (0.253,2.380) (0.193,1.827) -- (0.313,1.827) (0.193,2.380) -- (0.313,2.380);
\fill[tspec] (0.453,0) rectangle (0.783,1.728);
\draw[black!65,line width=0.45pt] (0.618,1.728) -- (0.618,1.728) (0.558,1.728) -- (0.677,1.728) (0.558,1.728) -- (0.677,1.728);
\fill[tmds] (0.818,0) rectangle (1.147,1.689);
\draw[black!65,line width=0.45pt] (0.983,1.689) -- (0.983,1.689) (0.923,1.689) -- (1.042,1.689) (0.923,1.689) -- (1.042,1.689);
\fill[tspring] (1.183,0) rectangle (1.513,1.701);
\draw[black!65,line width=0.45pt] (1.348,1.626) -- (1.348,1.877) (1.288,1.626) -- (1.408,1.626) (1.288,1.877) -- (1.408,1.877);
\node[below,align=center,font=\scriptsize] at (0.800,-0.05) {BB\\$[[18,4,4]]$};
\fill[halgray] (1.688,0) rectangle (2.018,3.196);
\draw[black!65,line width=0.45pt] (1.853,3.061) -- (1.853,3.323) (1.793,3.061) -- (1.913,3.061) (1.793,3.323) -- (1.913,3.323);
\fill[tspec] (2.053,0) rectangle (2.383,2.688);
\fill[tmds] (2.418,0) rectangle (2.748,2.674);
\fill[tspring] (2.783,0) rectangle (3.113,2.428);
\draw[black!65,line width=0.45pt] (2.948,2.267) -- (2.948,2.563) (2.888,2.267) -- (3.008,2.267) (2.888,2.563) -- (3.008,2.563);
\node[below,align=center,font=\scriptsize] at (2.400,-0.05) {BB\\$[[72,8,8]]$};
\fill[halgray] (3.288,0) rectangle (3.618,4.160);
\draw[black!65,line width=0.45pt] (3.453,3.840) -- (3.453,4.367) (3.393,3.840) -- (3.513,3.840) (3.393,4.367) -- (3.513,4.367);
\fill[tspec] (3.652,0) rectangle (3.982,2.966);
\fill[tmds] (4.018,0) rectangle (4.348,2.881);
\fill[tspring] (4.383,0) rectangle (4.713,2.942);
\draw[black!65,line width=0.45pt] (4.548,2.914) -- (4.548,2.961) (4.488,2.914) -- (4.607,2.914) (4.488,2.961) -- (4.607,2.961);
\node[below,align=center,font=\scriptsize] at (4.000,-0.05) {BB\\$[[144,12,12]]$};
\fill[halgray] (4.888,0) rectangle (5.218,1.407);
\draw[black!65,line width=0.45pt] (5.053,1.147) -- (5.053,1.692) (4.993,1.147) -- (5.112,1.147) (4.993,1.692) -- (5.112,1.692);
\fill[tspec] (5.253,0) rectangle (5.583,1.299);
\draw[black!65,line width=0.45pt] (5.418,1.299) -- (5.418,1.299) (5.358,1.299) -- (5.478,1.299) (5.358,1.299) -- (5.478,1.299);
\fill[tmds] (5.618,0) rectangle (5.948,1.454);
\draw[black!65,line width=0.45pt] (5.783,1.454) -- (5.783,1.454) (5.723,1.454) -- (5.843,1.454) (5.723,1.454) -- (5.843,1.454);
\fill[tspring] (5.982,0) rectangle (6.312,1.165);
\draw[black!65,line width=0.45pt] (6.147,1.098) -- (6.147,1.237) (6.088,1.098) -- (6.207,1.098) (6.088,1.237) -- (6.207,1.237);
\node[below,align=center,font=\scriptsize] at (5.600,-0.05) {Radial\\$[[24,2]]$};
\fill[halgray] (6.487,0) rectangle (6.817,1.617);
\draw[black!65,line width=0.45pt] (6.652,1.344) -- (6.652,1.900) (6.593,1.344) -- (6.712,1.344) (6.593,1.900) -- (6.712,1.900);
\fill[tspec] (6.853,0) rectangle (7.183,1.454);
\draw[black!65,line width=0.45pt] (7.018,1.454) -- (7.018,1.454) (6.958,1.454) -- (7.077,1.454) (6.958,1.454) -- (7.077,1.454);
\fill[tmds] (7.218,0) rectangle (7.548,1.233);
\draw[black!65,line width=0.45pt] (7.383,1.233) -- (7.383,1.233) (7.323,1.233) -- (7.442,1.233) (7.323,1.233) -- (7.442,1.233);
\fill[tspring] (7.582,0) rectangle (7.912,1.206);
\draw[black!65,line width=0.45pt] (7.747,1.145) -- (7.747,1.419) (7.688,1.145) -- (7.807,1.145) (7.688,1.419) -- (7.807,1.419);
\node[below,align=center,font=\scriptsize] at (7.200,-0.05) {Radial\\$[[32,2]]$};
\fill[halgray] (8.088,0) rectangle (8.417,2.922);
\draw[black!65,line width=0.45pt] (8.252,2.727) -- (8.252,3.125) (8.192,2.727) -- (8.312,2.727) (8.192,3.125) -- (8.312,3.125);
\fill[tspec] (8.453,0) rectangle (8.782,2.591);
\fill[tmds] (8.818,0) rectangle (9.147,2.589);
\fill[tspring] (9.183,0) rectangle (9.512,2.581);
\draw[black!65,line width=0.45pt] (9.348,2.566) -- (9.348,2.609) (9.287,2.566) -- (9.408,2.566) (9.287,2.609) -- (9.408,2.609);
\node[below,align=center,font=\scriptsize] at (8.800,-0.05) {Radial\\$[[54,8]]$};
\fill[halgray] (9.688,0) rectangle (10.017,2.266);
\draw[black!65,line width=0.45pt] (9.852,2.000) -- (9.852,2.474) (9.792,2.000) -- (9.912,2.000) (9.792,2.474) -- (9.912,2.474);
\fill[tspec] (10.053,0) rectangle (10.382,1.631);
\draw[black!65,line width=0.45pt] (10.217,1.631) -- (10.217,1.631) (10.157,1.631) -- (10.277,1.631) (10.157,1.631) -- (10.277,1.631);
\fill[tmds] (10.418,0) rectangle (10.747,1.899);
\draw[black!65,line width=0.45pt] (10.582,1.899) -- (10.582,1.899) (10.522,1.899) -- (10.643,1.899) (10.522,1.899) -- (10.643,1.899);
\fill[tspring] (10.783,0) rectangle (11.112,1.807);
\draw[black!65,line width=0.45pt] (10.947,1.670) -- (10.947,2.055) (10.887,1.670) -- (11.008,1.670) (10.887,2.055) -- (11.008,2.055);
\node[below,align=center,font=\scriptsize] at (10.400,-0.05) {qTanner\\$[[36,8,3]]$};
\fill[halgray] (11.287,0) rectangle (11.617,2.630);
\draw[black!65,line width=0.45pt] (11.452,2.441) -- (11.452,3.264) (11.392,2.441) -- (11.512,2.441) (11.392,3.264) -- (11.512,3.264);
\fill[tspec] (11.652,0) rectangle (11.982,1.910);
\draw[black!65,line width=0.45pt] (11.817,1.910) -- (11.817,1.910) (11.757,1.910) -- (11.877,1.910) (11.757,1.910) -- (11.877,1.910);
\fill[tmds] (12.018,0) rectangle (12.347,2.101);
\draw[black!65,line width=0.45pt] (12.182,2.101) -- (12.182,2.101) (12.122,2.101) -- (12.242,2.101) (12.122,2.101) -- (12.242,2.101);
\fill[tspring] (12.383,0) rectangle (12.712,2.129);
\draw[black!65,line width=0.45pt] (12.547,2.060) -- (12.547,2.200) (12.487,2.060) -- (12.607,2.060) (12.487,2.200) -- (12.607,2.200);
\node[below,align=center,font=\scriptsize] at (12.000,-0.05) {qTanner\\$n=54$};
\fill[halgray] (12.888,0) rectangle (13.217,3.002);
\draw[black!65,line width=0.45pt] (13.053,2.690) -- (13.053,3.170) (12.992,2.690) -- (13.113,2.690) (12.992,3.170) -- (13.113,3.170);
\fill[tspec] (13.253,0) rectangle (13.582,2.416);
\draw[black!65,line width=0.45pt] (13.418,2.416) -- (13.418,2.416) (13.357,2.416) -- (13.478,2.416) (13.357,2.416) -- (13.478,2.416);
\fill[tmds] (13.618,0) rectangle (13.947,2.224);
\draw[black!65,line width=0.45pt] (13.783,2.224) -- (13.783,2.224) (13.723,2.224) -- (13.843,2.224) (13.723,2.224) -- (13.843,2.224);
\fill[tspring] (13.983,0) rectangle (14.312,2.195);
\draw[black!65,line width=0.45pt] (14.148,2.127) -- (14.148,2.411) (14.088,2.127) -- (14.208,2.127) (14.088,2.411) -- (14.208,2.411);
\node[below,align=center,font=\scriptsize] at (13.600,-0.05) {qTanner\\$n=72$};
\fill[halgray] (0.20,4.850) rectangle (0.50,5.050); \node[right] at (0.55,4.950) {HAL default placement};
\fill[tspec] (4.30,4.850) rectangle (4.60,5.050); \node[right] at (4.65,4.950) {Spectral};
\fill[tmds] (7.00,4.850) rectangle (7.30,5.050); \node[right] at (7.35,4.950) {MDS};
\fill[tspring] (9.40,4.850) rectangle (9.70,5.050); \node[right] at (9.75,4.950) {Spring};
\end{tikzpicture}
  \caption{Excess HAL hardware complexity $C_{\mathrm{hw}}-1$ for nine qLDPC codes. A surface-code layout has $C_{\mathrm{hw}}=1$, and lower values are better. Gray bars show HAL's default placement. Teal bars show our placement with spectral embedding, MDS, and spring layout, routed by HAL. For HAL and spring layout, bars show the mean over seeds, and whiskers show the best and worst runs. Spectral embedding and MDS are deterministic.}
  \label{fig:hal}
\end{figure}

Figure~\ref{fig:hal} and Table~\ref{tab:hal} show the results. With spectral embedding or spring layout, our placement gives a lower $C_{\mathrm{hw}}$ than HAL's default placement for all nine codes. For the two randomized methods, HAL and spring layout, we average over the choice of seed. Relative to HAL's default mean, the excess $C_{\mathrm{hw}}-1$ drops by $8$--$29\%$ with spectral embedding and by $12$--$29\%$ with spring layout. MDS does better than HAL's default for eight codes and slightly worse for the radial $[[24,2]]$ code. We also compare to the best HAL run across the default, alternative, and toric layouts (Table~\ref{tab:hal-baselines}). The best spring-layout run has lower cost for every code and reduces the excess $C_{\mathrm{hw}}-1$ by $3$--$24\%$. Eight of the codes have at least six seeded runs per method. For each of them, a one-sided Mann--Whitney test comparing spring layout to HAL gives $p\le 0.0011$.

Most of the improvement comes from shorter couplers. Across the three embeddings, the average coupler length is shorter than with HAL's default placement by a factor of $1.7$--$3.4$. This agrees with the geometric costs calculated in the earlier sections, because placing Tanner-graph neighbors close together shortens the couplers between them. The numbers of tiers and through-silicon vias are less affected. They decrease for seven of the nine codes with spring layout and show no consistent trend with the other two embeddings.

We made two further checks. First, we changed the prefactor $c$ in $s = c\sqrt{|V_T|}$ from $0.5$ to $0.35$ or $0.7$. Across all nine codes and all three embeddings, our placement stays below HAL's default mean in $53$ of the $54$ cases (Table~\ref{tab:hal-sens}). Second, we tried a hybrid approach that keeps HAL's planar subgraph and tier assignment, but places this subgraph with our method instead of the Kamada--Kawai layout. For every code and embedding, the hybrid performs about as well as HAL's default and worse than our placement (Table~\ref{tab:hal-hybrid}). The improvement therefore comes from placing the whole Tanner graph.

We kept HAL's other settings, such as the grid aspect ratio, at the default values from \cite{HAL2025} for both placements. The score $C_{\mathrm{hw}}$ models the fabrication cost, and these results compare placements within that model.

\begin{table}[t]
\centering
\small
\caption{HAL hardware complexity $C_{\mathrm{hw}}$ (surface-code layout $=1$) with HAL's default placement and our placement for each continuous embedding algorithm. For HAL and spring layout we include both the mean and best score over the same random seeds; spectral embedding and MDS are deterministic single runs. ``Reduction'' is the percentage drop in excess $C_{\mathrm{hw}}-1$ relative to HAL's default mean, for spectral embedding, MDS, and the spring-layout mean. The last two columns give the average coupler length, which HAL normalizes by the shortest coupler on the first tier. BB codes are taken from HAL's generalized-toric parameter table.}
\label{tab:hal}
\begin{tabular}{lrccccccccc}
\toprule
 & & \multicolumn{2}{c}{HAL default} & & & \multicolumn{2}{c}{Spring} & & \multicolumn{2}{c}{Coupler length} \\
\cmidrule(lr){3-4}\cmidrule(lr){7-8}\cmidrule(lr){10-11}
Code & $|V_T|$ & mean & best & Spectral & MDS & mean & best & Reduction (\%) & HAL & S / M / Sp \\
\midrule
BB $[[18,4,4]]$ & 36 & 1.645 & 1.596 & 1.563 & 1.551 & 1.555 & 1.530 & $13$ / $15$ / $14$ & 5.4 & 2.5 / 2.6 / 2.5 \\
BB $[[72,8,8]]$ & 144 & 2.042 & 1.998 & 1.876 & 1.872 & 1.792 & 1.739 & $16$ / $16$ / $24$ & 10.4 & 5.4 / 5.3 / 4.2 \\
BB $[[144,12,12]]$ & 288 & 2.356 & 2.252 & 1.967 & 1.940 & 1.959 & 1.950 & $29$ / $31$ / $29$ & 14.6 & 6.0 / 5.5 / 5.8 \\
Radial $[[24,2]]$ & 48 & 1.459 & 1.374 & 1.424 & 1.474 & 1.380 & 1.358 & $8$ / $-3$ / $17$ & 3.8 & 1.8 / 2.2 / 1.7 \\
Radial $[[32,2]]$ & 64 & 1.527 & 1.438 & 1.474 & 1.402 & 1.393 & 1.373 & $10$ / $24$ / $25$ & 5.3 & 2.2 / 1.8 / 1.8 \\
Radial $[[54,8]]$ & 108 & 1.953 & 1.889 & 1.845 & 1.844 & 1.842 & 1.837 & $11$ / $11$ / $12$ & 9.0 & 5.0 / 5.4 / 4.5 \\
qTanner $[[36,8,3]]$ & 68 & 1.739 & 1.652 & 1.532 & 1.619 & 1.589 & 1.545 & $28$ / $16$ / $20$ & 7.6 & 2.2 / 3.0 / 3.0 \\
qTanner $n{=}54$ & 102 & 1.858 & 1.796 & 1.623 & 1.685 & 1.694 & 1.672 & $27$ / $20$ / $19$ & 8.9 & 3.6 / 4.0 / 4.3 \\
qTanner $n{=}72$ & 136 & 1.979 & 1.877 & 1.788 & 1.725 & 1.716 & 1.694 & $20$ / $26$ / $27$ & 10.2 & 4.3 / 4.9 / 4.0 \\
\bottomrule
\end{tabular}
\end{table}

\section{Conclusion}
\label{sec:conclusion}
We have developed a constructive pipeline for evaluating QECCs through the layout of their Tanner graphs. Continuous embedding provides a geometric view of a code and its Tanner-graph presentation; scale selection, grid assignment, and path routing turn that view into explicit paths on a target lattice. The resulting footprint, communication, and congestion proxies make it possible to compare layouts under common geometric assumptions, even when the underlying abstract code parameters are unchanged.

Across the codes studied, adding diagonal moves consistently reduces the total grid distance $W$, while adding a third dimension mainly relieves congestion and does not necessarily reduce $W$. The scale $s$ controls a trade-off between $W$ and congestion, and compressing a layout too much increases $W$ again. In our R2 HaPPY ablations, the continuous embedding and path-routing methods have more effect on the total cost than the choice between the Hungarian and greedy assignment algorithms. The ablation on nine qLDPC codes confirms that the continuous embedding matters more than the assignment algorithm. The embedding method matters as well: MDS, which preserves global distances, is less suited to keeping routed edges short, while spring layout can reduce communication at intermediate scales but raises congestion and varies more significantly with scale. Spectral embedding gives the most stable results in this comparison.

Our placement approach also helps under a hardware cost model. When our placement replaces HAL's default placement, HAL's router produces shorter couplers and a lower hardware complexity. This holds for all nine qLDPC codes we tested, for both spectral embedding and spring layout. A hardware-independent placement step can therefore be paired with platform-specific routers.

Some possible projects for the future would be to test our placement with routers for other platforms, such as neutral-atom arrays with their timing and movement constraints. One could also simulate the circuit-level logical performance of the resulting layouts and compare additional embedding, assignment, and path-routing algorithms. These steps would turn layout comparisons into predictions for specific QEC platforms.

\section*{Data availability}
The data underlying all figures and tables, including the raw HAL benchmark results, are available at \url{https://github.com/Yucheng-He/qec-embedding}.

\section*{Code availability}
The code for the layout pipeline, the ablation studies, and the HAL comparison is available at \url{https://github.com/Yucheng-He/qec-embedding}. HAL is publicly available at \url{https://github.com/EQuS/HAL}.

\section*{Use of AI tools}
AI coding assistants were used in developing parts of the simulation code. Claude (Anthropic) assisted with drafting parts of the text and the scripts for the HAL comparison and the qLDPC ablation. The authors designed the pipeline, directed all analyses, reviewed and verified all code, text, and numerical results, and take full responsibility for the content of this paper.

\section*{Competing interests}
The authors declare no competing interests.

\section*{Author contributions}
Y.H. developed the method, wrote the code, performed the simulations, and wrote the manuscript. T.A.B. supervised the project and revised the manuscript.

\begin{acknowledgments}
We thank ChunJun Cao, Kun Liu, Yongshan Ding, John Preskill, Qian Xu, and Min-Hsiu Hsieh for their helpful discussions. This work was supported in part by NSF Grants PHYS-2310794 and FET-2316713.
\end{acknowledgments}

\clearpage

\appendix

\section{Graph and Embedding Preliminaries}
\label{app:pipeline}

These appendices give the mathematical details of the default pipeline and of the ablation baselines. Throughout, the Tanner graph is denoted by
\[
G_T = (V_T,E_T),
\qquad
V_T = \{v_1,\dots,v_n\},
\]
so that \(n=|V_T|\) counts both data and check vertices. The target routing lattice is represented by the graph
\[
G_\Lambda=(\Lambda,\mathcal E_\Lambda).
\]
We write \(\mathbf x_i=p(v_i)\in\mathbb R^d\) for the continuous coordinates of vertex \(v_i\), and \(x\), \(y\) for lattice sites.

\subsection{Graph matrices and graph distance}

The adjacency matrix of the Tanner graph is
\[
A\in\{0,1\}^{n\times n},
\qquad
A_{ij}=1\iff (v_i,v_j)\in E_T,
\]
and the degree matrix is
\[
D=\operatorname{diag}(\deg_1,\dots,\deg_n),
\qquad
\deg_i = \sum_{j=1}^n A_{ij}.
\]
The unnormalized graph Laplacian is
\[
L = D-A.
\]
We use the unnormalized Laplacian throughout; normalized Laplacians would give alternative spectral embeddings. The input to MDS is the graph-distance matrix
\[
\Delta_T = \big(d_T(v_i,v_j)\big)_{i,j=1}^n,
\]
where \(d_T\) is the shortest-path distance on the Tanner graph.

Because \(d_T\) is infinite between disconnected components, distance-based embeddings require a connected Tanner graph. All Tanner graphs considered in this paper are connected, and our implementation checks this before embedding.

The two distances \(d_T\) and \(d_{\mathrm{grid}}\) should not be confused. The Tanner-graph distance \(d_T\) is defined on the graph \(G_T\) alone, while the lattice distance \(d_{\mathrm{grid}}\) is a distance between sites of \(G_\Lambda\), and is evaluated only after grid assignment. A routed path can be longer than \(d_{\mathrm{grid}}\) between its endpoints if it takes detours.

\subsection{Continuous embeddings}

A continuous embedding is a map
\[
p:V_T\to\mathbb R^d .
\]
Each embedding method optimizes its own objective on \(p\), such as the strain for MDS or the quadratic form \(\mathcal Q(p)\) for spectral embedding, but we do not compare methods by these objectives. All metrics (\(F\), \(W_{\rm grid}\), \(\mathrm{LB}\), \(\mathrm{UB}\), and \(\mathrm{ParLB}\)) are computed on the routed layout obtained after scale selection, grid assignment, and path routing.

\subsection{Scale sweep}
\label{app:scale_sweep}

The overall scale of \(p\) is arbitrary. For each candidate scale \(s>0\), we form the scaled positions
\[
\hat p_s(v)=s\,p(v),
\]
solve the grid-assignment problem, route all Tanner-graph edges, and evaluate
\[
F(s),\qquad W_{\rm grid}(s),\qquad \mathrm{LB}(s),\qquad \mathrm{UB}(s),\qquad \mathrm{ParLB}(s),
\]
together with the total cost \(J(s)\) of Eq.~\eqref{eq:scale_obj_new}. The selected scale \(s^\ast\) is the candidate value that minimizes \(J(s)\). In the R2 HaPPY ablation studies of Sec.~\ref{sec:ablation}, the candidate scales are \(s=1.0,1.5,\dots,21.0\), but other ranges could also be used. The qLDPC ablation uses the scales given in Table~\ref{tab:ablation-qldpc}. Figures that illustrate the effect of \(s\), such as Fig.~\ref{assort}, also show non-optimal scales.

Different embedding methods return coordinates on different numerical scales, so before the sweep through different values of $s$, we center each graph-aware embedding (spectral, MDS, or spring) and rescale it so that its largest coordinate magnitude equals one. The type-stratified baseline, a graph-blind initializer that places the vertices in rows by type (Appendix~\ref{app:type_stratified}), is rescaled as described there. With this normalization, the same value of \(s\) gives layouts of comparable extent for different methods.

\section{Multidimensional Scaling}
\label{app:mds}

MDS is the distance-based method among our three continuous embeddings. We use it to visualize the global geometry of a Tanner graph (Sec.~\ref{sec:continuous_init}) and in the within-stage substitution (Sec.~\ref{sec:ablation}).

\subsection{Classical MDS}

We use classical (Torgerson) MDS~\cite{Torgerson1952MDS}, which computes the coordinates in closed form. Let \(\Delta_T^{(2)}\) be the matrix of squared graph distances, with entries \(d_T(v_i,v_j)^2\), and let \(H = I - \tfrac1n\mathbf 1\mathbf 1^T\) be the centering matrix. Double centering, that is, multiplying by \(H\) on both sides, which subtracts the row and column means, gives the matrix
\begin{equation}
B = -\tfrac12\,H\,\Delta_T^{(2)}\,H .
\end{equation}
If the graph distances were exactly Euclidean distances between points \(\mathbf x_i\) with zero mean, \(B\) would be their Gram matrix, \(B_{ij} = \mathbf x_i\cdot\mathbf x_j\). Classical MDS therefore takes the \(d\) largest eigenvalues \(\gamma_1 \ge \dots \ge \gamma_d\) of \(B\) with orthonormal eigenvectors \(w_1,\dots,w_d\), and sets the \(m\)th coordinate of vertex \(v_i\) to \(\sqrt{\gamma_m}\,(w_m)_i\); negative eigenvalues, which appear because graph distances are generally not Euclidean, are set to zero. The resulting coordinates minimize the strain \(\|B-PP^T\|_F^2\) over all \(n\times d\) coordinate matrices \(P\), whose rows are \(\mathbf x_i^T\), where \(\|\cdot\|_F\) is the Frobenius norm.

The strain is not the normalized stress \(\sigma(p)\) of Sec.~\ref{sec:continuous_init}. We use the stress only to diagnose how well the embedding preserves graph distances.

\subsection{Why MDS emphasizes global distances}

The stress sums over all \(\binom{n}{2}\) vertex pairs with equal weight, so a pair of adjacent vertices counts no more than a pair at opposite ends of the graph; the same is true of the strain. MDS will therefore stretch some edges if doing so improves the fit to long distances. This is useful for visualizing global geometry, but it is less aligned with the total cost, whose communication term depends only on the lengths of Tanner-graph edges. This mismatch is consistent with the higher routed communication of MDS observed in Sec.~\ref{sec:ablation}.

\section{Spectral Embedding}
\label{app:spectral}

Spectral embedding is our default continuous embedding. Its coordinates are determined up to the sign of each eigenvector and the choice of basis within degenerate eigenspaces.

\subsection{Laplacian quadratic form}

Let \(\xi\in\mathbb R^n\) assign a real number \(\xi_i\) to each Tanner-graph vertex \(v_i\). Then
\begin{equation}
\xi^TL\xi = \xi^T(D-A)\xi = \sum_{i=1}^n \deg_i\, \xi_i^2 - \sum_{i,j=1}^n A_{ij}\xi_i\xi_j,
\end{equation}
and by the symmetry of \(A\),
\begin{equation}
\xi^TL\xi = \frac{1}{2}\sum_{i,j=1}^n A_{ij}(\xi_i-\xi_j)^2 = \sum_{(v_i,v_j)\in E_T}(\xi_i-\xi_j)^2 .
\end{equation}
The quadratic form therefore penalizes separating adjacent vertices. Applying it to each of the \(d\) coordinate columns of an embedding matrix \(\Theta\in\mathbb R^{n\times d}\) and summing gives \(\operatorname{tr}(\Theta^TL\Theta) = \mathcal Q(p)\), the edge-based objective of Sec.~\ref{sec:continuous_init}.

\subsection{Rayleigh--Ritz characterization}

Because \(L\mathbf 1=0\), the constant vector lies in the nullspace of \(L\). For a connected graph, this gives the trivial eigenpair
\[
\lambda_1=0,
\qquad
\phi_1\propto \mathbf 1,
\]
and all other eigenvalues are positive, \(0 < \lambda_2 \le \lambda_3 \le \dots \le \lambda_n\).

The first nontrivial coordinate solves
\[
\min_{\xi\in\mathbb R^n} \xi^TL\xi
\quad\text{subject to}\quad
\|\xi\|_2=1,
\qquad
\xi\perp \mathbf 1.
\]
By the Rayleigh--Ritz theorem, the minimizer is an eigenvector \(\phi_2\) associated with the smallest nonzero eigenvalue \(\lambda_2\), and the minimum value is \(\lambda_2\). Intuitively, \(\phi_2\) is the smoothest nonconstant function on the graph: it changes as little as possible along edges.

The \(d\)-dimensional spectral embedding uses the first \(d\) nontrivial eigenvectors,
\[
\Theta = [\phi_2\;\phi_3\;\cdots\;\phi_{d+1}]\in\mathbb R^{n\times d},
\]
so that vertex \(v_i\) receives the coordinates
\begin{equation}
p(v_i)=\big(\phi_2(i),\phi_3(i),\dots,\phi_{d+1}(i)\big).
\end{equation}
By the same variational argument, this \(\Theta\) minimizes \(\operatorname{tr}(\Theta^TL\Theta)\) subject to \(\Theta^T\Theta = I\) and \(\Theta^T\mathbf 1 = 0\), with minimum value \(\lambda_2 + \dots + \lambda_{d+1}\); these are the normalization and orthogonality constraints referred to in Sec.~\ref{sec:continuous_init}. We compute the embedding with the NetworkX \texttt{spectral\_layout} routine~\cite{NetworkX2008}, which uses the unnormalized Laplacian.

\subsection{Why spectral embedding highlights local structure}

Because the objective penalizes coordinate differences only across edges, densely connected subgraphs are assigned similar coordinates, while sparse cuts and bottlenecks appear as directions along which the eigenvectors change rapidly. This makes the modular and layered structure easy to see. The unit normalization of the eigenvectors fixes an arbitrary overall scale (the individual coordinates are of order \(n^{-1/2}\)), which is then adjusted by the rescaling and scale sweep steps described in Appendix~\ref{app:scale_sweep}.

\section{Spring Layout}
\label{app:spring}

Spring (force-directed) layout treats the Tanner-graph vertices as particles in a mechanical system that is relaxed iteratively, with attraction along Tanner-graph edges and repulsion between vertices. The conventions below apply to the within-stage substitution in Sec.~\ref{sec:ablation}.

\subsection{Fruchterman--Reingold forces}

We use the Fruchterman--Reingold variant of spring layout~\cite{Fruchterman1991}. For a distance \(r = \|\mathbf x_i - \mathbf x_j\|_2\) between vertices \(i\) and \(j\), the magnitudes of the attractive and repulsive forces are
\begin{equation}
f_a(r)=\frac{r^2}{k}
\qquad\text{(along edges \((v_i,v_j)\in E_T\))},
\end{equation}
\begin{equation}
f_r(r)=\frac{k^2}{r}
\qquad\text{(between every pair \(i\neq j\))},
\end{equation}
where \(k > 0\) sets the natural spring length: an isolated pair of adjacent vertices is in equilibrium at \(r = k\). Fruchterman and Reingold choose \(k = C \sqrt{\text{area}/|V_T|}\), where area is the area of the frame in which the layout is drawn and \(C\) is a constant of order one, so that each vertex occupies an area of order \(k^2\). At each iteration, every vertex moves along the net force acting on it, with a step size capped by a temperature that decreases over the iterations, so that the system anneals toward a stable configuration.

Unlike spectral embedding and classical MDS, this procedure has no closed-form solution; its output depends on \(k\), the temperature schedule, the number of iterations and the random initialization.

\subsection{Implementation conventions}

We generate spring-layout coordinates with the NetworkX \texttt{spring\_layout} routine~\cite{NetworkX2008}, which uses \(k = 1/\sqrt{|V_T|}\) (a frame of unit area and \(C = 1\)), a temperature that decreases linearly over the iterations, and random initial positions. We use its default of \(50\) iterations and random seed \(42\). The resulting coordinates then go through the same normalization, scale sweep, grid assignment, and path routing as those of MDS and spectral embedding.

\subsection{Why spring layout can succeed at intermediate scales}

In the Fruchterman--Reingold model, attraction along an edge grows quadratically with its length, whereas repulsion decays inversely with distance. Densely connected groups of vertices therefore contract into tight clusters. Within a cluster edges are short, which lowers the communication cost at intermediate scales (Fig.~\ref{fig:ablation-within-stage}). However, the same compactness makes many routed paths pass through a small number of lattice sites, raising congestion. This is a plausible mechanism for the trade-off seen in Sec.~\ref{sec:ablation}.

\section{Grid Assignment and the Hungarian Algorithm}
\label{app:grid_assignment}

\subsection{From scaled positions to an assignment problem}
\label{app:assignment_routing_conventions}

At a fixed scale \(s\), grid assignment solves the linear assignment problem (LAP) of Eq.~\eqref{eq:lap}, with costs \(c_{v\ell}\) and assignment variables \(x_{v\ell}\) for \(v\in V_T\) and \(\ell\in\mathcal S\). Its cost matrix has size \(|V_T|\times|\mathcal S|\) and is generally rectangular, since the set of candidate sites \(\mathcal S\) usually contains more sites than there are vertices. Adding \(|\mathcal S|-|V_T|\) dummy vertices, whose cost is zero for every site, turns it into a square matrix of size \(N = |\mathcal S|\): every optimal assignment of the square problem restricts to an optimal assignment of the original one, and the sites given to dummy vertices are simply left empty.

\subsection{The Hungarian method}
\label{app:hungarian}

For a square problem, relaxing \(x_{v\ell}\in\{0,1\}\) to \(x_{v\ell}\ge 0\) defines a linear program whose constraints \(\sum_\ell x_{v\ell}=1\) and \(\sum_v x_{v\ell}=1\) define the Birkhoff polytope. Its vertices are exactly the permutation matrices, so the linear program has an optimal solution that is a one-to-one assignment, and the relaxation is tight. The dual linear program is
\begin{equation}
\max_{\mu,\nu}\ \sum_{v}\mu_v + \sum_{\ell}\nu_\ell
\quad\text{subject to}\quad
\mu_v + \nu_\ell\le c_{v\ell} \quad\text{for all } v,\ell,
\end{equation}
where the dual variables \(\mu_v\) and \(\nu_\ell\) are called potentials. For feasible potentials, the reduced costs \(\bar c_{v\ell} = c_{v\ell} - \mu_v - \nu_\ell\) are nonnegative. If a complete one-to-one assignment uses only pairs with \(\bar c_{v\ell} = 0\), its total cost equals \(\sum_v \mu_v + \sum_\ell \nu_\ell\); by weak duality, no assignment can cost less, so this assignment is optimal.

The Hungarian method~\cite{Kuhn1955Hungarian,Munkres1957Assignment} searches for such an assignment. It maintains feasible potentials and a partial assignment that uses only zero-reduced-cost pairs. To add an unassigned vertex, it searches for a path that starts at this vertex, ends at a free site, and uses only zero-reduced-cost pairs, alternating between unassigned and assigned pairs; flipping the pairs along such a path assigns one more vertex. If no such path exists, it raises \(\mu_v\) for every vertex reached by the search and lowers \(\nu_\ell\) for every site reached, both by the smallest reduced cost from a reached vertex to an unreached site. This keeps the potentials feasible, leaves the reduced costs of the pairs in the current search tree unchanged, and creates at least one new zero-reduced-cost pair so the search can continue. Each vertex is added after at most \(N\) such updates, and the full assignment is found in \(O(N^3)\) time. In practice, we use the SciPy routine \texttt{linear\_sum\_assignment}, which implements a shortest-augmenting-path version of this method and handles rectangular cost matrices directly.

\subsection{Application to grid assignment}
\label{app:candidate_sites}

In grid assignment, rows correspond to Tanner-graph vertices and columns to the candidate sites in \(\mathcal S\), with cost \(c_{v\ell} = \|\hat p_s(v) - \ell\|_*\) in the norm of Eq.~\eqref{eqn:Euclidean1inf}. In the implementation, the cost is evaluated from the rounded scaled position \(p_0(v) = \operatorname{round}(\hat p_s(v))\), i.e., \(c_{v\ell} = \|p_0(v) - \ell\|_* + \varepsilon_{v\ell}\), where \(\varepsilon_{v\ell}\) is a small deterministic perturbation of order \(10^{-9}\) times the column index of \(\ell\), included only to break ties. By the triangle inequality, this differs from the ideal cost \(\|\hat p_s(v) - \ell\|_*\) by at most \(\|\hat p_s(v) - p_0(v)\|_* + |\varepsilon_{v\ell}|\), that is, by at most \(1/2\) in the octilinear (\(\ell_\infty\)) case or \(d/2\) in the rectilinear (\(\ell_1\)) case, plus the perturbation. This bound does not depend on \(s\). The Hungarian algorithm exactly minimizes the implemented cost.

Likewise, the candidate set \(\Gamma_R(v)\) is built around \(p_0(v)\) rather than \(\hat p_s(v)\), with \(R=1\). It contains the lattice sites within distance 1 of \(p_0(v)\) in the matched norm. The candidate sets are used only to build \(\mathcal S\): in the LAP, any vertex may be assigned to any site in \(\mathcal S\), not only to sites in its own set \(\Gamma_R(v)\). The LAP is therefore feasible whenever \(|\mathcal S| \geq  |V_T|\). If \(|\mathcal S| < |V_T|\), we add the lattice sites nearest to the scaled positions until \(|\mathcal S| = |V_T|\).

\section{Path Routing}
\label{app:routing}

After grid assignment, each Tanner-graph edge \((u,v)\in E_T\) is routed as a lattice path \(\pi_{uv}\) between \(\tilde p(u)\) and \(\tilde p(v)\). The default router builds a shortest path in the lattice metric \(d_{\rm grid}\) directly from the endpoint coordinates.

\subsection{Move sets}

In \(d\)-dimensional rectilinear geometry, the allowed moves are the axis-aligned unit steps
\[
\mathcal M_{\rm rect}=\{\pm e_m: m=1,\dots,d\},
\]
where \(e_m\) is the \(m\)th unit vector. In octilinear geometry, the allowed moves are the nonzero sign vectors
\[
\mathcal M_{\rm oct}=\{-1,0,1\}^d\setminus\{0\}.
\]
In 2D, these are the four axis-aligned and four diagonal steps. Every allowed move has unit cost, regardless of its Euclidean length.

\subsection{Shortest path distances}

Let \(x\) and \(y\) be lattice sites with coordinate offsets \(\delta_m = |x_m - y_m|\), \(m = 1,\dots,d\).

In rectilinear geometry, each move changes exactly one coordinate by one unit, so it reduces \(\sum_m \delta_m = \|x - y\|_1\) by at most one. Every path from \(x\) to \(y\) therefore has at least \(\|x - y\|_1\) steps, and correcting the offsets one axis at a time attains this bound:
\begin{equation}
d_{\rm grid}(x,y) = \|x - y\|_1 .
\end{equation}

In octilinear geometry, each move changes every coordinate by at most one unit, so it reduces \(\max_m \delta_m = \|x - y\|_\infty\) by at most one, and every path has at least \(\|x - y\|_\infty\) steps. Conversely, moving at each step along the sign vector of the current offset reduces every nonzero offset by one unit and reaches \(y\) in exactly \(\max_m \delta_m\) steps. Hence,
\begin{equation}
d_{\rm grid}(x,y) = \|x - y\|_\infty .
\end{equation}

\subsection{Canonical shortest paths}

Shortest paths are generally not unique, and the router fixes one deterministically. In rectilinear geometry, it corrects the coordinate offsets one axis at a time in a fixed order. In octilinear geometry, it moves along the sign vector of the current offset, as in the construction above; as individual offsets reach zero, the moves involve fewer axes, and the final moves are axis-aligned. Because every edge is routed along a shortest path, \(\operatorname{len}(\pi_{uv}) = d_{\rm grid}(\tilde p(u),\tilde p(v))\) for every edge, and this choice minimizes the communication proxy \(\mathrm{UB}\). Congestion-aware detours are used only in the A* baseline below.

\section{Ablation and Baseline Algorithms}
\label{app:baselines}

The cross-stage ablation in Fig.~\ref{fig:ablation-J} replaces one stage of the default pipeline (spectral embedding, Hungarian algorithm, and shortest-path routing) at a time, on the R2 HaPPY code. All ablation runs use 2D octilinear geometry, the weights \((w_F,w_{\rm UB},w_{\rm Par})=(1,0.02,0.5)\), and the scales listed in Appendix~\ref{app:scale_sweep}. The qLDPC ablation in Table~\ref{tab:ablation-qldpc} uses the same baselines and settings, except for its range of scales.

\subsection{Type-stratified initialization}
\label{app:type_stratified}

The type-stratified initializer is a deliberately graph-blind baseline: it uses only the vertex type (data qubit, \(X\)-check, or \(Z\)-check) and ignores Tanner-graph connectivity. It places the three types of vertices on three parallel rows, with data vertices at \(y = 0\), \(X\)-check vertices at \(y = 1\), and \(Z\)-check vertices at \(y = -1\). Within each row, vertices are ordered by their labels from left to right along \(x\) at unit spacing, centered at \(x = 0\). The layout is then centered and rescaled so that the larger of its two coordinate ranges equals one.

\subsection{Greedy assignment}
\label{app:greedy}

Greedy assignment starts from the same scaled positions \(\hat p_s(v)\) as the default (Hungarian) grid assignment, but replaces the global one-to-one matching with a sequential local search:
\begin{enumerate}
\item Sort the vertices lexicographically by \((\hat p_{s,1}(v),\hat p_{s,2}(v),\operatorname{str}(v))\), where \(\hat p_{s,1}\) and \(\hat p_{s,2}\) are the first two coordinates of \(\hat p_s\) and \(\operatorname{str}(v)\) is a deterministic string label used to break ties.
\item In that order, round the scaled position of each vertex to a lattice site \(p_0(v)=\operatorname{round}(\hat p_s(v))\).
\item If \(p_0(v)\) is free, assign \(v\) there. Otherwise, search outward in square shells of increasing radius around \(p_0(v)\), and assign \(v\) to the closest free site, in the norm \(\|\cdot\|_*\), within the first shell that contains one, breaking ties by taking the site with the smallest first coordinate and then the smallest second coordinate.
\end{enumerate}
Unlike the Hungarian assignment, an early vertex never gives up its site to a later one, so the result depends on the processing order and need not minimize the total displacement. Because this baseline uses the same scaled positions, lattice, and norm as the default grid assignment, it isolates the effect of replacing the global matching by a sequential local search. It is labeled ``w/o Hungarian algorithm'' in Fig.~\ref{fig:ablation-J}.

\subsection{Congestion-aware A* routing}
\label{app:astar}

For a source site \(a\) and a target site \(b\), A* explores lattice sites in order of increasing \(g(x)+h(x)\)~\cite{HartNilssonRaphael1968AStar}, where \(g(x)\) is the cost of the best path found so far from \(a\) to \(x\), and \(h(x)\) is an estimate of the remaining cost to \(b\):
\begin{equation}
h_{\rm rect}(x) = \|x - b\|_1,
\qquad
h_{\rm oct}(x) = \|x - b\|_\infty.
\end{equation}
For a move from site \(x\) to an adjacent site \(y\), the congestion-aware cost is
\begin{equation}
c(x,y) = 1 + \kappa\,\eta_{\{x,y\}} + \beta\,\mathbf 1_{\rm bend},
\end{equation}
where \(\mathbf 1_{\rm bend} = 1\) if the move changes direction relative to the previous move and \(0\) otherwise, and \(\eta_{\{x,y\}}\) is the number of previously routed paths that use the lattice edge \(\{x,y\}\),
\begin{equation}
\eta_e = \sum_{\pi\in\mathcal P} \mathbf 1[e\in \pi],
\end{equation}
with \(\mathcal P\) the set of paths routed so far. The parameter \(\kappa \ge 0\) sets the congestion penalty, and \(\beta \ge 0\) is an optional bend penalty. Every move costs at least one, and \(h(x)\) is the minimum number of moves from \(x\) to \(b\), so neither heuristic overestimates the remaining cost (in A* terminology, both are admissible).

In the A* baseline shown in Fig.~\ref{fig:ablation-J}, \(\kappa=1\) and \(\beta=2\). The edges are routed one at a time in a single pass, in lexicographic order of their endpoint labels, without rerouting earlier paths, and each search is restricted to the bounding box of the placed vertices enlarged by five sites on each side. The search keeps one state per lattice site rather than per site and incoming direction, so with \(\beta > 0\) it is a heuristic: the returned path need not exactly minimize the total move cost. A* detours around heavily used lattice edges, which lowers congestion (\(\mathrm{ParLB}\)) but increases the footprint and path length.

\subsection{Combined baseline}
\label{app:without_all}

The combined ``w/o all'' baseline replaces all three default stages at once: it uses type-stratified initialization, greedy assignment, and congestion-aware A* routing.

\begin{table}[h]
\centering
\caption{Scale sensitivity of our placement in HAL: $C_{\mathrm{hw}}$ at $s=c\sqrt{|V_T|}$ for $c=0.35$, $0.5$, and $0.7$, compared with the mean for HAL's default placement. Spring layout uses seed $0$ at every scale.}
\label{tab:hal-sens}
\begin{tabular}{llcccc}
\toprule
Code & Embedding & $c=0.35$ & $c=0.5$ & $c=0.7$ & HAL default (mean)\\
\midrule
BB $[[18,4,4]]$ & spectral & 1.589 & 1.563 & 1.552 & 1.645 \\
BB $[[18,4,4]]$ & MDS & 1.571 & 1.551 & 1.543 & 1.645 \\
BB $[[18,4,4]]$ & spring & 1.544 & 1.560 & 1.603 & 1.645 \\
BB $[[72,8,8]]$ & spectral & 1.855 & 1.876 & 1.924 & 2.042 \\
BB $[[72,8,8]]$ & MDS & 1.850 & 1.872 & 1.926 & 2.042 \\
BB $[[72,8,8]]$ & spring & 1.868 & 1.808 & 1.738 & 2.042 \\
BB $[[144,12,12]]$ & spectral & 2.124 & 1.967 & 1.974 & 2.356 \\
BB $[[144,12,12]]$ & MDS & 2.139 & 1.940 & 1.867 & 2.356 \\
BB $[[144,12,12]]$ & spring & 2.141 & 1.966 & 1.919 & 2.356 \\
Radial $[[24,2]]$ & spectral & 1.385 & 1.424 & 1.469 & 1.459 \\
Radial $[[24,2]]$ & MDS & 1.413 & 1.474 & 1.407 & 1.459 \\
Radial $[[24,2]]$ & spring & 1.396 & 1.358 & 1.371 & 1.459 \\
Radial $[[32,2]]$ & spectral & 1.504 & 1.474 & 1.475 & 1.527 \\
Radial $[[32,2]]$ & MDS & 1.401 & 1.402 & 1.469 & 1.527 \\
Radial $[[32,2]]$ & spring & 1.419 & 1.390 & 1.349 & 1.527 \\
Radial $[[54,8]]$ & spectral & 1.874 & 1.845 & 1.851 & 1.953 \\
Radial $[[54,8]]$ & MDS & 1.876 & 1.844 & 1.845 & 1.953 \\
Radial $[[54,8]]$ & spring & 1.849 & 1.839 & 1.764 & 1.953 \\
qTanner $[[36,8,3]]$ & spectral & 1.571 & 1.532 & 1.631 & 1.739 \\
qTanner $[[36,8,3]]$ & MDS & 1.582 & 1.619 & 1.570 & 1.739 \\
qTanner $[[36,8,3]]$ & spring & 1.580 & 1.569 & 1.628 & 1.739 \\
qTanner $n{=}54$ & spectral & 1.751 & 1.623 & 1.674 & 1.858 \\
qTanner $n{=}54$ & MDS & 1.829 & 1.685 & 1.669 & 1.858 \\
qTanner $n{=}54$ & spring & 1.711 & 1.680 & 1.650 & 1.858 \\
qTanner $n{=}72$ & spectral & 1.701 & 1.788 & 1.756 & 1.979 \\
qTanner $n{=}72$ & MDS & 1.744 & 1.725 & 1.711 & 1.979 \\
qTanner $n{=}72$ & spring & 1.801 & 1.786 & 1.752 & 1.979 \\
\bottomrule
\end{tabular}
\end{table}

\section{Details of the HAL comparison}
\label{app:hal}

We ran HAL~\cite{HAL2025} without modification and left all its settings at their default values. The routing grid has $500\times500$ cells, and every node and routed edge keeps a margin of one cell. Each coupler is limited to at most $10$ bump-bond transitions and a maximum length of $1000$ cells. HAL routes edges in order of increasing length and computes the hardware complexity with its default normalization. We generated the Tanner graphs with HAL's code constructors so that both placements act on identical graphs. Specifically, we used its generalized-toric parameter table for the BB codes, a fixed random seed for the radial codes, and its parity-check files for the quantum Tanner codes.

\emph{HAL default.} HAL first computes a randomized community layout and uses it to order the edges. Following this order, it extracts a planar subgraph for the first tier. It then draws this subgraph with a Kamada--Kawai layout scaled by $\lceil\sqrt{|V_T|}\rceil$, rounds the vertex positions to distinct integer points, and removes empty rows and columns. All other edges are routed on higher tiers. To compare with HAL's best result rather than only its default, we also replaced the community layout with HAL's Kamada--Kawai, spectral, and spring layouts, one run each, and ran HAL's toric placement for the three BB codes. Table~\ref{tab:hal-baselines} compares our placement with the best of all these HAL runs.

\begin{table}[h]
\centering
\small
\caption{Hybrid variant (HAL tiering with our placement applied to the planar subgraph) compared with HAL's default placement and with our placement, for each continuous embedding. Entries are $C_{\mathrm{hw}}$. HAL default and hybrid: means over HAL seeds. Our placement: single runs for spectral embedding and MDS, mean over seeds for spring layout.}
\label{tab:hal-hybrid}
\begin{tabular}{lccccccc}
\toprule
 & & \multicolumn{3}{c}{Hybrid} & \multicolumn{3}{c}{Our placement} \\
\cmidrule(lr){3-5}\cmidrule(lr){6-8}
Code & HAL default & Spectral & MDS & Spring & Spectral & MDS & Spring \\
\midrule
BB $[[18,4,4]]$ & 1.645 & 1.634 & 1.634 & 1.675 & 1.563 & 1.551 & 1.555 \\
BB $[[72,8,8]]$ & 2.042 & 2.033 & 2.107 & 2.062 & 1.876 & 1.872 & 1.792 \\
BB $[[144,12,12]]$ & 2.356 & 2.433 & 2.426 & 2.358 & 1.967 & 1.940 & 1.959 \\
Radial $[[24,2]]$ & 1.459 & 1.548 & 1.556 & 1.529 & 1.424 & 1.474 & 1.380 \\
Radial $[[32,2]]$ & 1.527 & 1.529 & 1.664 & 1.523 & 1.474 & 1.402 & 1.393 \\
Radial $[[54,8]]$ & 1.953 & 2.028 & 1.960 & 1.920 & 1.845 & 1.844 & 1.842 \\
qTanner $[[36,8,3]]$ & 1.739 & 1.730 & 1.784 & 1.745 & 1.532 & 1.619 & 1.589 \\
qTanner $n{=}54$ & 1.858 & 1.812 & 1.911 & 1.807 & 1.623 & 1.685 & 1.694 \\
qTanner $n{=}72$ & 1.979 & 1.982 & 1.922 & 1.982 & 1.788 & 1.725 & 1.716 \\
\bottomrule
\end{tabular}
\end{table}

\begin{table}[h]
\centering
\small
\caption{HAL with different initial layouts, compared with our placement. HAL columns: best run with the default community layout (random seeds), single runs with HAL's Kamada--Kawai, spectral, and spring initial layouts, HAL's toric placement (BB codes only), and the best of all these HAL runs. Our columns: spectral embedding and MDS (deterministic) and the best spring-layout run. Entries are $C_{\mathrm{hw}}$.}
\label{tab:hal-baselines}
\begin{tabular}{lccccccccc}
\toprule
 & \multicolumn{6}{c}{HAL initial layout} & \multicolumn{3}{c}{Ours} \\
\cmidrule(lr){2-7}\cmidrule(lr){8-10}
Code & community & K.--K. & spectral & spring & toric & best & Spectral & MDS & Spring \\
\midrule
BB $[[18,4,4]]$ & 1.596 & 1.604 & 1.565 & 1.617 & 1.545 & 1.545 & 1.563 & 1.551 & 1.530 \\
BB $[[72,8,8]]$ & 1.998 & 2.082 & 1.934 & 2.059 & 2.268 & 1.934 & 1.876 & 1.872 & 1.739 \\
BB $[[144,12,12]]$ & 2.252 & 2.344 & 2.338 & 2.286 & 2.243 & 2.243 & 1.967 & 1.940 & 1.950 \\
Radial $[[24,2]]$ & 1.374 & 1.471 & 1.628 & 1.572 & -- & 1.374 & 1.424 & 1.474 & 1.358 \\
Radial $[[32,2]]$ & 1.438 & 1.465 & 1.568 & 1.436 & -- & 1.436 & 1.474 & 1.402 & 1.373 \\
Radial $[[54,8]]$ & 1.889 & 2.066 & 1.973 & 2.054 & -- & 1.889 & 1.845 & 1.844 & 1.837 \\
qTanner $[[36,8,3]]$ & 1.652 & 1.734 & 1.641 & 1.687 & -- & 1.641 & 1.532 & 1.619 & 1.545 \\
qTanner $n{=}54$ & 1.796 & 1.877 & 1.880 & 1.782 & -- & 1.782 & 1.623 & 1.685 & 1.672 \\
qTanner $n{=}72$ & 1.877 & 1.990 & 1.997 & 1.916 & -- & 1.877 & 1.788 & 1.725 & 1.694 \\
\bottomrule
\end{tabular}
\end{table}

\emph{Our placement.} We compute a spectral, MDS, or spring embedding of the full Tanner graph, scale it by $s=0.5\sqrt{|V_T|}$, and assign the vertices to distinct sites of a rectilinear lattice with the Hungarian algorithm of Sec.~\ref{subsec:grid_assignment}. The integer coordinates are then passed to HAL as its first-tier placement. In this mode, HAL keeps nearest-neighbor edges on the first tier, tries to route the remaining edges there as well, and moves any unrouted edges to higher tiers.

\emph{Comparison with the best HAL run.} Table~\ref{tab:hal-baselines} lists all HAL baselines. For every code, the best spring-layout run beats the best HAL run and reduces the excess $C_{\mathrm{hw}}-1$ by $3$--$24\%$. Spectral embedding and MDS give one layout each. Spectral embedding beats the best HAL run on every code except the $[[18,4,4]]$ BB code and the radial $[[24,2]]$ and $[[32,2]]$ codes. MDS beats it on every code except the $[[18,4,4]]$ BB code and the radial $[[24,2]]$ code. For the $[[18,4,4]]$ BB code, the best HAL result comes from the toric placement, which follows the lattice structure of the code. For the $[[144,12,12]]$ BB code, each method has three runs, and every spring-layout run lies below every HAL run. This is the strongest outcome that a Mann--Whitney test can register for this sample size ($p=0.05$).

\emph{Scale sensitivity.} Table~\ref{tab:hal-sens} lists $C_{\mathrm{hw}}$ at $s=c\sqrt{|V_T|}$ for three values of the prefactor $c$. Over the $54$ cases with $c=0.35$ or $0.7$, $C_{\mathrm{hw}}$ changes by at most $0.20$ relative to $c=0.5$. Only spectral embedding at $c=0.7$ for the radial $[[24,2]]$ code lies above HAL's default mean, by $0.01$. At $s=1.0\sqrt{|V_T|}$, our placement of the radial $[[32,2]]$ code had no nearest-neighbor edges. Because HAL's benchmark does not support that case, we do not report it.

\emph{Hybrid variant.} The hybrid approach keeps HAL's community layout, planar subgraph, and tier assignment, but places this subgraph with our method instead of the Kamada--Kawai layout. We ran the hybrid with all three embeddings. We used three seeds of HAL's community layout for the five smallest codes, two seeds for the next three codes, and one seed for the $[[144,12,12]]$ BB code. For every code and embedding, the hybrid's mean $C_{\mathrm{hw}}$ lies within $0.14$ of HAL's default mean and is $0.05$--$0.49$ higher than our placement with the same embedding (see Table~\ref{tab:hal-hybrid}).

\section{Interpretation and Limitations of Proxy Metrics}
\label{app:metric_limitations}

The quantities \(F\), \(W_{\rm grid}\), \(\mathrm{LB}\), \(\mathrm{UB}\), and \(\mathrm{ParLB}\) are layout-level proxies for comparing Tanner-graph layouts and routing geometries without a full hardware compilation.

\begin{itemize}
\item \(W_{\rm grid}\) (Eq.~\eqref{eq:W}) is the total endpoint-to-endpoint lattice distance of the assignment, before any routing detours.
\item \(\mathrm{LB}\) (Eq.~\eqref{eq:LB}) depends only on endpoint distances; it is a lower-bound proxy, not an exact gate or schedule count.
\item \(\mathrm{UB}\) (Eq.~\eqref{eq:UB}) is evaluated along the routed paths. Under shortest-path routing it equals \(2W_{\rm grid}\); under detouring routers it also counts the extra path length.
\item \(\mathrm{ParLB}\) (Eq.~\eqref{eq:ParLB}) is a lower-bound proxy for serialization based on the maximum edge and node loads, not an exact syndrome-extraction circuit depth.
\item \(F\) (Eq.~\eqref{eq:footprint}) counts occupied lattice sites, whether they hold qubits, couplers, relays, or transport paths; it is not the number of physical data qubits.
\end{itemize}

These proxies do not model measurement and ancilla-reset timing, decoder latency and feedforward, platform-specific gate times and error rates, crosstalk, or detailed shuttling and coupler protocols. Those effects belong to a downstream, hardware-specific compiler.

\bibliographystyle{phaip}
\bibliography{reference}

\end{document}